\documentclass[a4, 12pt]{article}
\usepackage[pdftex]{graphicx}
\usepackage{amsmath,amssymb,amsfonts,amscd,amsfonts,epsf,epsfig,color,url}
\usepackage{enumitem}
\usepackage{enumitem}

\allowdisplaybreaks[4]

\date{}
\begin{document}

\begin{flushright} 

\end{flushright} 

\vspace{0.1cm}

\begin{center}
  {\LARGE
  
Color Confinement and Bose-Einstein Condensation

}
\vspace{5mm}

\end{center}
\vspace{0.1cm}
\vspace{0.1cm}
\begin{center}

Masanori Hanada$^b$, Hidehiko Shimada$^e$ and Nico Wintergerst$^c$

\end{center}
\vspace{0.3cm}

\begin{center}

$^b${\it Department of Mathematics, University of Surrey, Guildford, Surrey, GU2 7XH, UK}\\
\vspace{3mm}
$^e${\it Mathematical and Theoretical Physics Unit, Okinawa Institute of Science and Technology,}\\
{\it 1919-1 Tancha, Onna-son, Okinawa 904-0495 Japan}\\
\vspace{1mm}
{\it Yukawa Institute for Theoretical Physics, Kyoto University\\
Kitashirakawa Oiwakecho, Sakyo-ku, Kyoto 606-8502 Japan}\\
\vspace{3mm}
$^c${\it The Niels Bohr Institute, University of Copenhagen,}\\
{\it  Blegdamsvej 17, 2100 Copenhagen \O, Denmark} 

\end{center}

\vspace{1.5cm}

\begin{center}
  {\bf Abstract}
\end{center}

We propose a unified description of two important phenomena:
color confinement in large-$N$ gauge theory, and Bose-Einstein condensation (BEC).  
We focus on the confinement/deconfinement transition characterized by the increase of the entropy from $N^0$ to $N^2$, which persists in the weak coupling region.
Indistinguishability associated with the symmetry group --- SU($N$) or O($N$) in gauge theory, 
and S$_N$ permutations in the system of identical bosons --- is crucial 
 for the formation of the condensed (confined) phase.
We relate standard criteria,
based on off-diagonal long range order (ODLRO) for BEC
and the Polyakov loop for gauge theory.
The constant offset of the distribution of the phases of the Polyakov loop corresponds to ODLRO, 
and gives the order parameter for the partially-(de)confined phase at finite coupling.
We demonstrate this explicitly for several quantum mechanical systems (i.e., theories at small or zero spatial volume) at weak coupling, 
and argue that this mechanism extends to large volume and/or strong coupling.
This viewpoint may have implications for confinement at finite $N$, 
and for quantum gravity via gauge/gravity duality.

\vspace{2cm}
\begin{center}
MH would like to dedicate this paper to Keitaro Nagata. 
\end{center}

\newpage
\tableofcontents

\section{Introduction}
\hspace{0.51cm}

In this paper we point out 
a hitherto unnoticed
connection between two important phase transitions: 
Bose-Einstein condensation (BEC)~\cite{einstein1924quantentheorie} 
and the confinement/deconfinement transition
\cite{Polyakov:1978vu,Susskind:1979up}
in large-$N$ gauge theories.  

Throughout this paper, we adopt a characterization of 
confinement and deconfinement by the increase of energy and entropy from 
order $N^0$ to $N^2$ (for fields in the adjoint representation), 
or to $N^1$ (for fields in the fundamental representation).
There are two important motivations to consider the confinement/deconfinement transition in this sense. 
First, 
in the context of 
gauge/gravity duality,
a detailed understanding of this transition may provide important insight on 
how information about the spacetime geometry 
is encoded on the gauge theory side. 
The confined and deconfined phases
are considered to describe the vacuum and black hole geometry on the
gravity side, respectively~\cite{Witten:1998zw}.
Second, this characterization may lead to a new way of understanding the more traditional
`dynamical' confinement \cite{Wilson:1974sk} 
defined
in terms of the existence of a linear potential between probe quarks. 
This approach was pioneered in Refs. \cite{Sundborg:1999ue,Aharony:2003sx}
in which gauge theories defined on a spatial sphere are considered.
It was shown that 
the confinement characterized by the $N$-dependence of the free energy
carries over to weak or even vanishing coupling, and 
small or vanishing spatial volume, where the system and the transition 
can be described analytically.\footnote{For asymptotically free theories, the small volume limit
and the weak coupling limit are identical.}
For finite interaction and large volume, 
the confinement defined by the two characterizations are expected to coincide, 
as argued e.g. in~\cite{Hanada:2014noa}.\footnote{
In order to extract lessons applicable to the strong coupling regime 
from weak-coupling calculations,
the crucial assumption is that 
the strong- and weak-coupling regions are smoothly connected without any phase boundary.
Whether such a phase boundary is absent or not is model dependent.
For some models, the comparisons with strong coupling results 
obtained via holography and/or lattice simulation gave credible support to 
the absence of the phase boundary; 
see e.g. Refs.~\cite{Aharony:2005bq,Aharony:2005ew}.
For an elaboration on this point, see the discussion section.
}

\begin{figure}[htbp]
\begin{center}
\scalebox{0.2}{
\includegraphics{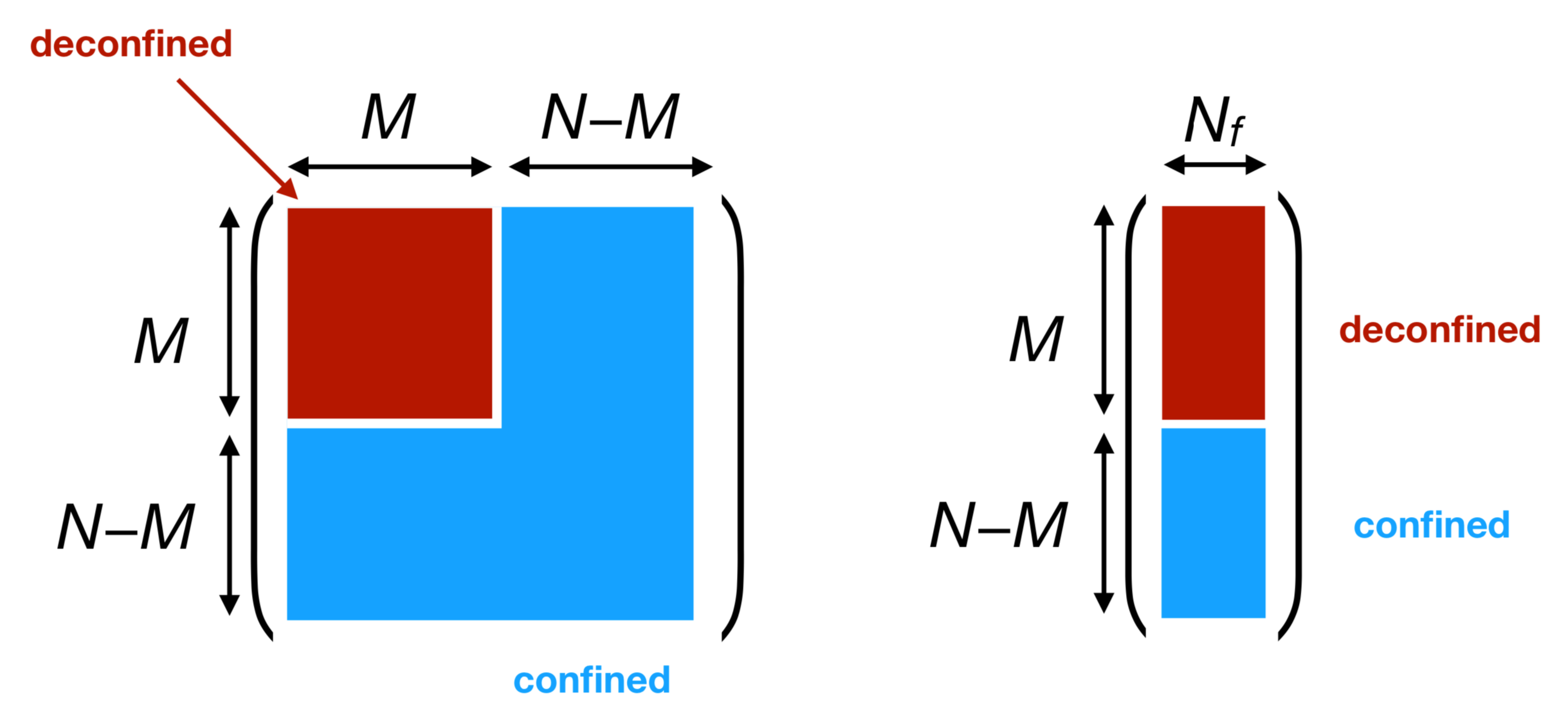}}
\end{center}
\caption{Partial confinement in the gauge sector with adjoint matter (left) and vector matter (right). 
The elements shown in blue are confined,
whereas the elements shown in red are deconfined.
These figures are taken from Ref.~\cite{Hanada:2019czd}.
Partial confinement can be defined in a gauge-invariant manner. 
For details, see Sec.~\ref{sec:Yang-Mills}, around Eq.~\eqref{eq:gauge_theory_symmetrize}.
}\label{fig:partial-deconfinement}
\end{figure}

The purpose of this paper is twofold.
First, we 
obtain a better
 understanding of the mechanism behind 
the confinement/deconfinement transition in the weakly-coupled regime
by exploiting the connection to the physics of BEC. 
As concrete and solvable examples, we consider SU($N$) Yang-Mills on a small three sphere, the gauged O($N$) vector model on a small two-sphere, 
and a system of $N$ identical bosons. While all those examples are essentially quantum mechanics rather than quantum field theory,\footnote{
For quantum field theory compactified on a small sphere where nonzero-momentum modes can be integrated out, the effective theory of zero-momentum modes is essentially quantum mechanics. }
the underlying mechanism we identify is applicable to quantum field theory as well. 
Second, we consider the problem of extending the understanding at weak-coupling
 to the strongly-coupling regime. 
Here, the analogy to BEC provides us with new insight on
the order parameters which may be used to discriminate the confined and the deconfined
phases for all values of the coupling.

The key concept underlying the connection between confinement and BEC is the 
idea of {\it partial confinement} recently introduced in 
Refs.~\cite{Hanada:2016pwv,Berenstein:2018lrm,Hanada:2018zxn,Hanada:2019czd,Hanada:2019kue}.
In the partially-confined phase\footnote{
Whether to call this phase partial confinement or
partial deconfinement is of course purely a matter of taste.
In this paper, we prefer to use the term partial {\it confinement},
since it parallels the term Bose-Einstein condensation. } of Yang-Mills theory, 
only the degrees of freedom associated with the $M\times M$ submatrix 
(red region in Fig.~\ref{fig:partial-deconfinement}) are 
deconfined, whereas 
the remaining degrees of freedom (blue region in Fig.~\ref{fig:partial-deconfinement}) are confined.\footnote{
How partial confinement is reconciled with gauge invariance is explained in detail in Ref.~\cite{Hanada:2019czd}. See also Sec.~\ref{sec:Yang-Mills}. }
In this sense, the confined and deconfined sectors coexist in the space of colors,
and thermodynamic quantities, for example 
entropy, in the partially-confined phase can be understood 
as a sum of two terms associated with the two components, 
i.e.  the confined and the deconfined sectors.
In Refs.~\cite{Hanada:2019czd,Hanada:2019kue}, the existence of a 
partially-confined phase was demonstrated for several weakly-coupled theories
by explicitly counting the states contributing to
thermodynamic quantities. 
It was further shown in Refs.~\cite{Hanada:2019czd,Hanada:2019kue} that, for weakly coupled theories, 
the confined degrees of freedom are in their ground state.
Thus, as we will elaborate in Sec.~\ref{sec:weak-coupling}, 
the confinement/deconfinement transition has the following 
characteristic features:
\begin{enumerate}
\item
The transition occurs even in the weak-coupling limit which can capture 
important parts of the physics of the transition in strongly-coupled systems.

\item
In the (partially) confined phase, 
a large fraction of the degrees of freedom falls into the ground state (the confined sector).

\item
The ground state and the excited states (the deconfined sector) coexist 
and thermodynamics can be understood from the point of view of 
a system with these two components.

\end{enumerate}
Furthermore, as we will show in Sec.~\ref{sec:underlying_mechanism}, 
partial confinement also has the following feature:
\begin{enumerate}[resume]
\item
Positive interference due to the gauge symmetry is the key mechanism
of confinement. 
\end{enumerate}

The crucial element of our proposal is that precise counterparts of the above features
exist for BEC. Namely: 
\begin{enumerate}
\item
The transition occurs even in the weak-coupling limit (the ideal Bose gas)
which can capture important parts of the physics of the transition in strongly-coupled systems.

\item
In the condensed phase, 
a large fraction of particles fall into the ground state (the Bose-Einstein condensate).   

\item
The system consists of particles in the ground state 
and excited states. The thermodynamic properties can be
understood from the point of view of 
a system with these two components.

\item
Positive interference due to the permutation symmetry is 
the key mechanism of condensation. 
\end{enumerate}

A standard example of BEC with non-vanishing interactions is the superfluidity of $^4$He. 
The confined and deconfined sectors in Yang-Mills theories correspond to 
super- and normal-fluid components, respectively.  
One may find it surprising that the weak-coupling calculation of 4d Yang-Mills
\cite{Sundborg:1999ue,Aharony:2003sx} captures 
the essence of strongly-coupled dynamics 
obtained by lattice simulation or holography. 
From our new point of view, this parallels the fact that a good part of 
the characteristic features of superfluidity in $^4$He, 
which is interacting via the van der Waals force, 
can be understood starting with the free theory, 
as first pointed out by F.~London \cite{1938Natur.141..643L}.

The connection between confinement and BEC becomes particularly transparent in 
a model that is almost tailor-made for this purpose:
 the gauged O($N$) vector model. 
In Sec.~\ref{sec:weak-coupling}, we show how
confinement in this model \cite{Shenker:2011zf} 
is related to 
confinement in Yang-Mills, and to BEC in 
a system of $N$ identical bosons, 
focussing on the essential features 1.-3. listed above.

Once appreciating the analogy explained in Sec.~\ref{sec:weak-coupling}, 
it is straightforward to 
uncover the common mechanism behind 
confinement and BEC: the indistinguishability (or equivalently, the redundancy) of  states
due to gauge symmetry or permutation symmetry
leads to a parametrically large 
enhancement of the ground state, known as \emph{Bose enhancement} or 
\emph{positive interference of the ground-state wave function}.  
We will explain this mechanism in Sec.~\ref{sec:underlying_mechanism},
and find a precise, quantitative characterization of 
this interference effect in Sec.~\ref{sec:Pol-vs-ODLRO}. 

As is well-known, the finite temperature system is described by the Euclidean path integral
with the compactified temporal direction.
The Polyakov loop $P$ is defined by the gauge covariant 
path ordered exponential $P= {\rm P} e^{\int A_0 dt}$ along a closed path extended in the
temporal direction.
(The trace of $P$ is also called Polyakov loop.)
Since the path ordered exponential is a unitary matrix, its eigenvalues,
which are of course gauge invariant, are phase factors of 
the form $e^{i \theta_j}$ $(j=1, \cdots, N)$. Later we will use the
density of these phases $\rho(\theta)$ in the large-$N$ limit. 
$P$ and $\rho$ can depend on the spatial position of the temporal loop.
When used as the order parameter for the confinement/deconfinement transition, 
usually the spatial average is considered.
In this paper we consider the Polyakov loop in the fundamental representation.
There is a deep connection to a class of phase transtions characterized 
by the behavior of the Polyakov loop,
first advocated by Gross, Witten \cite{Gross:1980he} and Wadia \cite{Wadia:2012fr,Wadia:1980cp} (the GWW transition). 
That is, the description of a system of identical bosons as 
a theory with S$_N$ gauge symmetry permits a 
straightforward definition of a Polyakov loop, and we show 
that the formation of a BEC can also be interpreted as a GWW transition.\footnote{
Here we define the `GWW transition' by the disappearance of the gap in the distribution of Polyakov loop phases. 
Other details of the phase transition, including the order of transition, 
depend on model-specific details such as the dimension and matter content.  
Note that the original context of the GWW model (arising from the 2D Yang-Mills theory)
is not relevant in our discussion.
}
Therefore,
both confinement and BEC are characterized by the change of the distribution of the phases of Polyakov loop. 
We will further show that this characterization of the transition based on the
Polyakov loop is closely related to 
the more traditional characterization based on off-diagonal 
long range order (ODLRO) \cite{PhysRev.104.576,RevModPhys.34.694}.
This argument readily generalizes to Yang-Mills models as well.
In particular, for an ideal Bose gas, 
we prove explicitly that ODLRO and the criteria based on the
Polyakov loop are equivalent.

In Sec.~\ref{sec:discussions} we discuss possible applications to 
quantum gravity via holography and confinement in finite-$N$ theories. 

Throughout the paper, we set the Planck constant $\hbar$ and Boltzmann constant $k_{\rm B}$ to be unity.

\section{The correspondence in the weak-coupling limit}\label{sec:weak-coupling}
\hspace{0.51cm}
Let us commence our analysis at zero coupling.
First, we provide the necessary background on partial confinement in Yang-Mills theory in Sec.~\ref{sec:Yang-Mills}.  
The gauged O($N$) vector model is particularly well suited to 
establish the connection between confinement in Yang-Mills and BEC, 
and we will introduce it in Sec.~\ref{sec:O(N)-vector-model}. 
We will close the section by explaining BEC in the system of identical bosons
in Sec.~\ref{sec:BEC-identical-bosons}. 
These three examples form the basis of the connection between confinement and BEC. 
The common underlying mechanism
 will be explained in Sec.~\ref{sec:underlying_mechanism}. 

Note that Sec.~\ref{sec:Yang-Mills} and Sec.~\ref{sec:O(N)-vector-model} are based on Ref.~\cite{Hanada:2019czd}, and Sec.~\ref{sec:BEC-identical-bosons} explains well-known 
established results.
We will present the known results in such a way that the unknown connection is revealed.\footnote{
Some readers may find it helpful to refresh their memory about the 
basic feature of BEC explained in Sec.~2.3 
before reading Sec.~2.1
in order to appreciate the close analogy between the two phenomena.
}
\subsection{Yang-Mills theory at weak coupling}\label{sec:Yang-Mills}
\hspace{0.51cm}
As a typical example of the confinement/deconfinement transition in large-$N$ Yang-Mills theory,
we consider finite temperature pure Yang-Mills theory defined on the space 
S$^3$ with the gauge group SU($N$).
We consider the free limit
of pure Yang-Mills theory,
which is solvable analytically \cite{Sundborg:1999ue,Aharony:2003sx}.  
As we have mentioned in the introduction, and as we will explicitly demonstrate, 
{\it partial confinement takes place even in the free limit
and the free theory captures important features of the 
confinement/deconfinement transition.}
See e.g.~Refs.~\cite{Aharony:2005bq,Aharony:2005ew} regarding the resemblance
between weak- and strong-coupling regions.
Most of the dynamical degrees of freedom
can be integrated out, since they become massive
due to the compactness and the curvature of S$^3$.
In this way, an effective action for the phases of the Polyakov loop is obtained.\footnote{
More precisely, the gauge conditions $\partial^iA_i=0$ and $\partial_t\alpha=0$, 
where $\alpha=\int d^3x_{{\rm S}^3}A_0(x)\times\frac{1}{{\rm Vol}_{{\rm S}^3}}$, are imposed, 
and the Polyakov loop is defined by $P=e^{i\beta\alpha}$.  
For more details, see Ref.~\cite{Aharony:2003sx}.
}  
This effective action can be solved by using standard matrix-model techniques and
the results can be naturally explained in terms of partial confinement, as we now review.

Let us start with a precise definition of the partially-confined sector.
In a generic partially-confined state, $M\times M$ degrees of freedom
are excited. The remaining degrees of freedom are
in the ground state, as shown in Fig.~\ref{fig:partial-deconfinement}.
The partially-confined states in the Hilbert space of the theory 
can be constructed in a manifestly gauge-invariant manner. 
First, we consider a trivial embedding of SU($M$) into SU($N$), as the upper-left block
(Fig.~\ref{fig:partial-deconfinement}). 
All SU($M$)-invariant energy eigenstates $|E;{\rm SU}(M)\rangle$ are then obtained
by exciting the oscillators associated with the $M\times M$ submatrix components while
respecting the $M\times M$ part of the Gauss law constraints. 
When doing this, we keep all oscillators associated with the remaining 
$N^2-M^2$ elements (i.e. the elements shown in blue in Fig.~\ref{fig:partial-deconfinement}) in their ground states. 
The states thus prepared are invariant under 
an SU($M$)$\times$SU($N-M$) subgroup of the SU($N$) gauge symmetry.
Finally, to construct the fully SU($N$)-invariant state
$|E\rangle_{\rm inv}$ with the same energy,
we act with all possible gauge transformation on this state and take the superposition:
\begin{eqnarray}
|E\rangle_{\rm inv}
=
{\cal N}^{-1/2}\int dU\ {\cal U}\left(|E;{\rm SU}(M)\rangle\right). 
\label{eq:gauge_theory_symmetrize}
\end{eqnarray}  
Here ${\cal U}$ stands for the SU($N$) transformation acting on the states in the Hilbert space, which corresponds to the group element $U\in{\rm SU}(N)$.\footnote{
When the states are expressed by acting the creation operators $\hat{a}^\dagger_{\mu,ij}$ to the Fock vacuum, 
$\left({\cal U}\hat{a}^\dagger_{\mu}\right)_{ij}
=
\sum_{k,l=1}^NU_{ik}\hat{a}^\dagger_{\mu,kl}U^{-1}_{lj}$. 
}
The integral is taken over SU($N$).  
The normalization factor ${\cal N}^{-1/2}$ ensures unit normalization of $|E\rangle_{\rm inv}$.
The mapping from the SU($M$)$\times$SU($N-M$)-invariant states to the SU($N$)-invariant states is one-to-one.
This allows one to straightforwardly count all such states and explicitly show that they dominate the thermodynamics~\cite{Hanada:2019czd}. 
In this way, {\it a large fraction of the degrees of freedom falls into the ground state
}(the confined sector), and the {\it confined sector and deconfined sector coexist}. 

\begin{figure}[htbp]
\begin{center}
\scalebox{0.2}{
\includegraphics{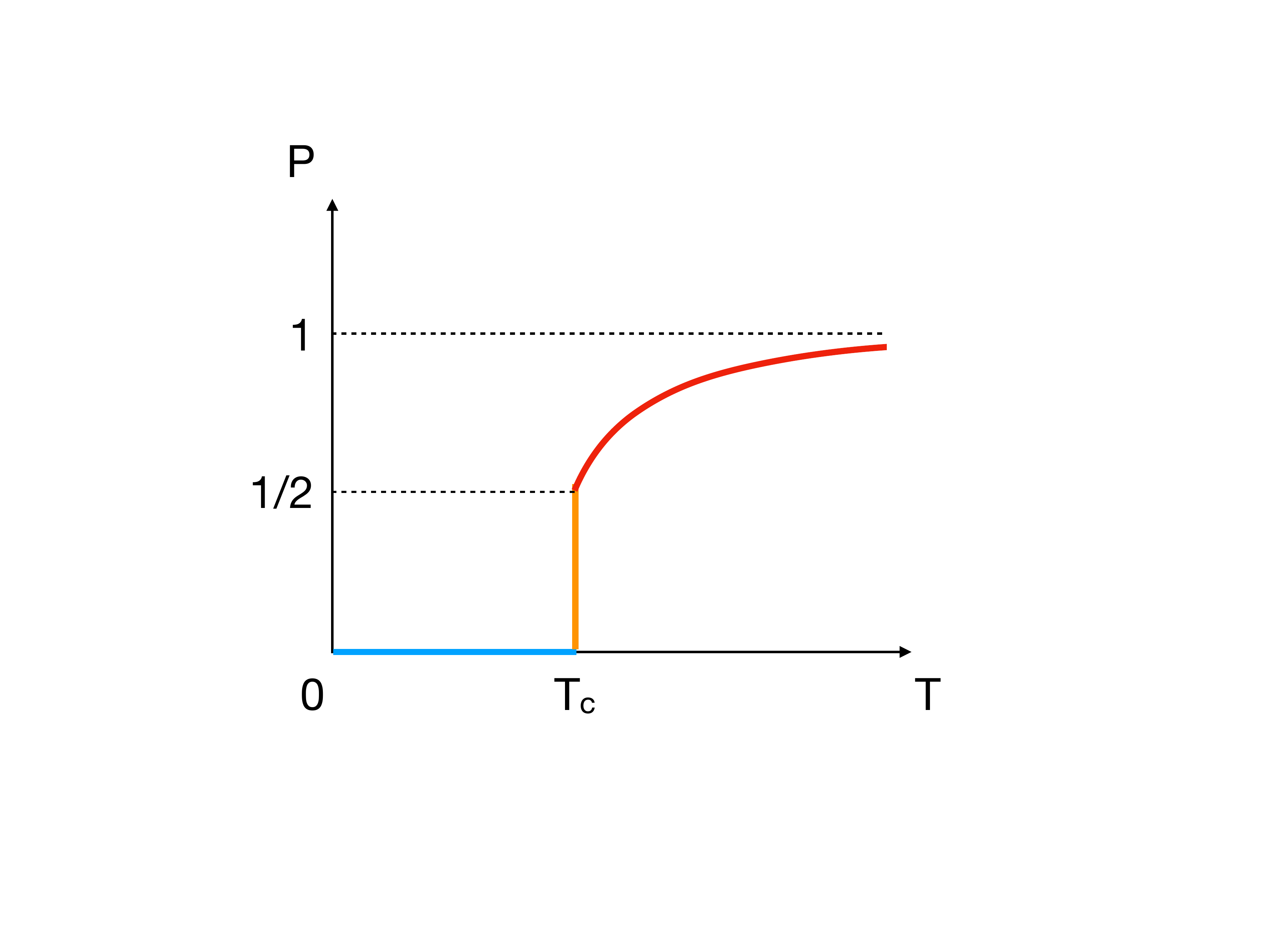}}
\scalebox{0.2}{
\includegraphics{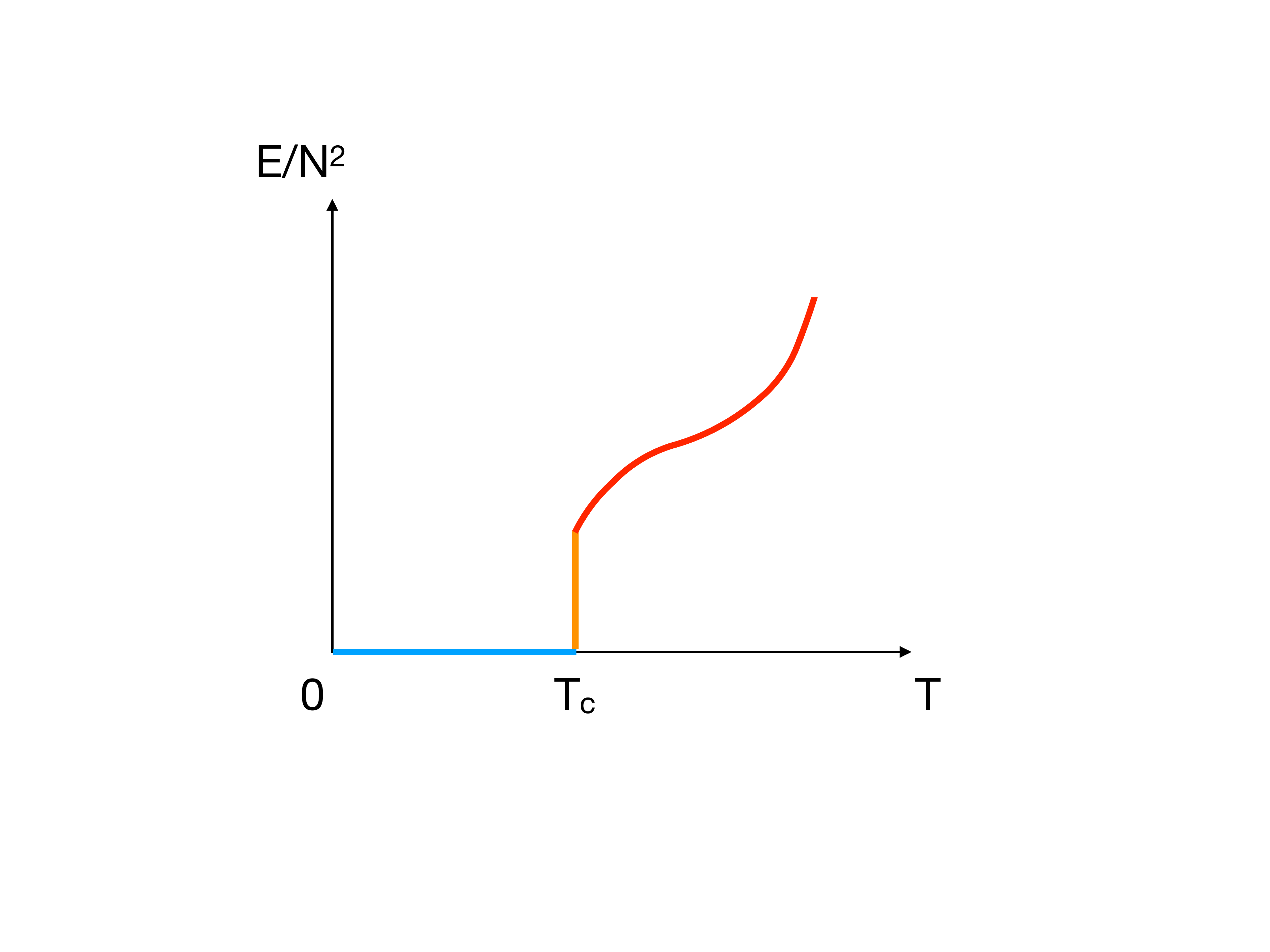}}
\scalebox{0.2}{
\includegraphics{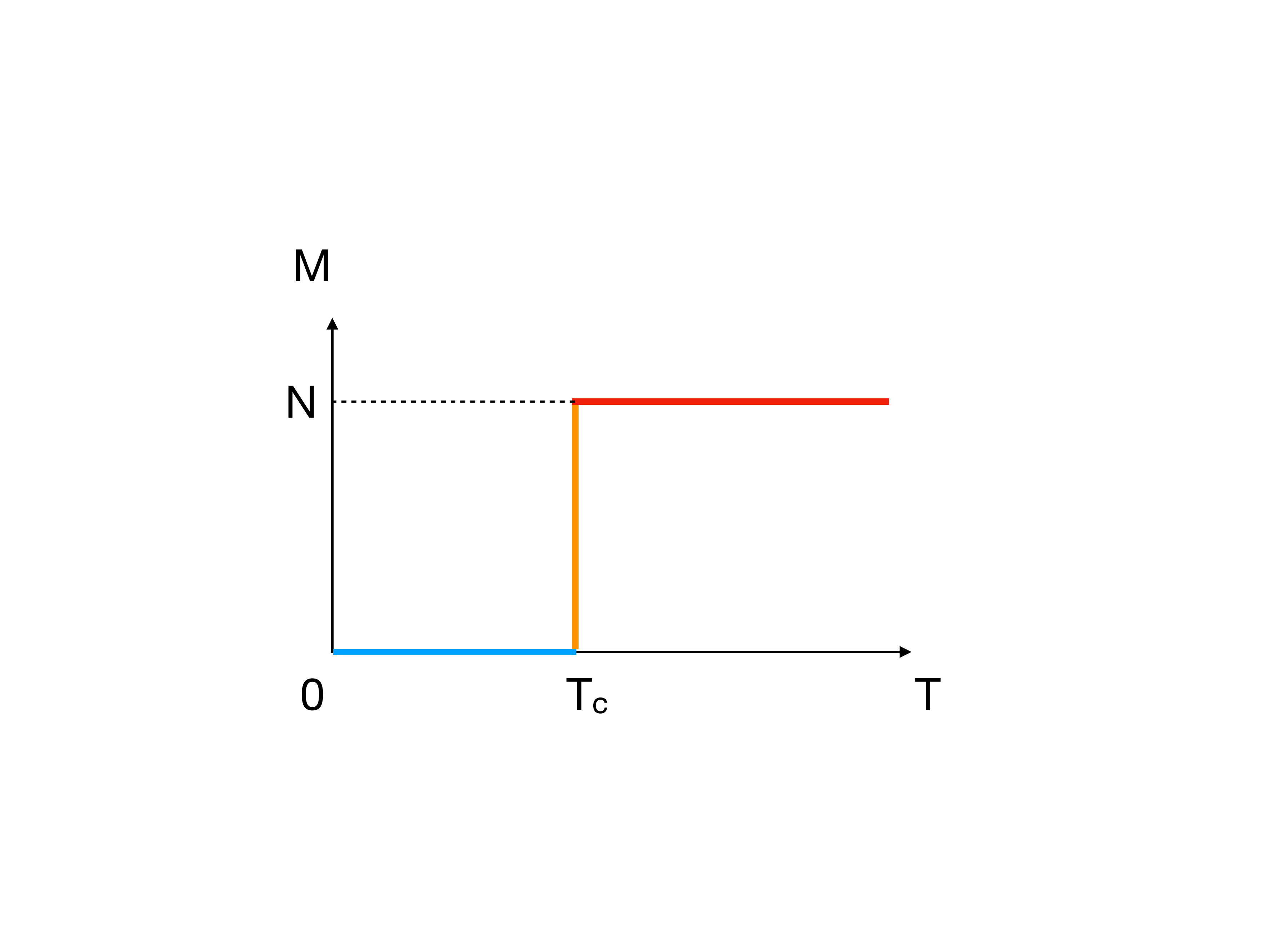}}
\end{center}
\caption{
Qualitative features 
of the phases of the free Yang-Mills on S$^3$
\cite{Hanada:2018zxn,Hanada:2019czd}. 
The Polyakov loop $P$, energy $E$ and the size of the deconfined sector $M$ 
in weakly-coupled 4d YM on S$^3$
are shown as functions of temperature $T$. 
The blue, orange and red lines are completely confined, partially confined
(equivalently partially deconfined) and 
completely deconfined phases, respectively. 
}\label{fig:partial-deconfinement-free-YM}
\end{figure}

\begin{figure}[htbp]
\begin{center}
\scalebox{0.4}{
\includegraphics{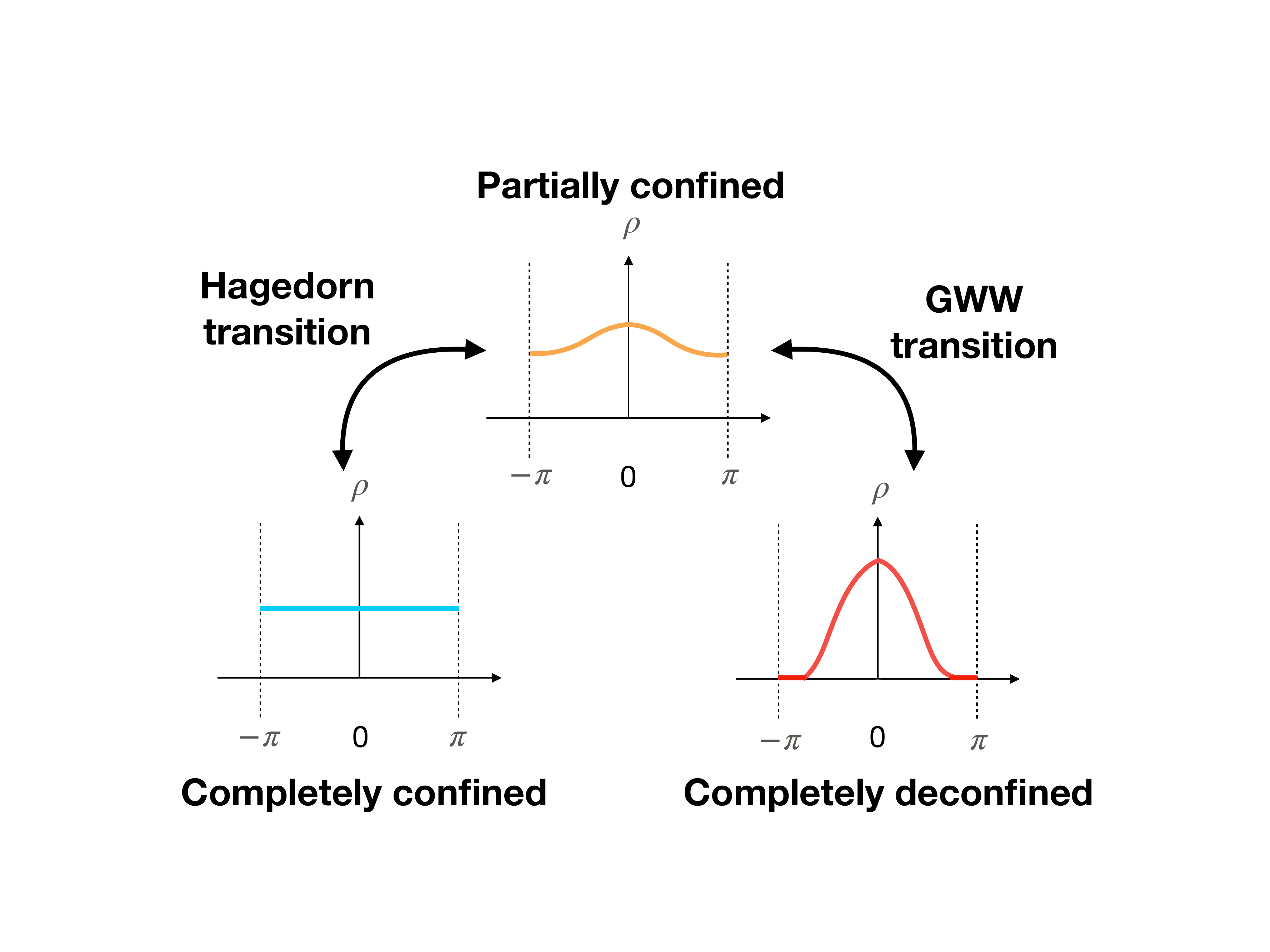}}
\end{center}
\caption{The distribution of the phases of Polyakov loop in the
completely confined, partially confined and completely deconfined phases. 
This figure is taken from Ref.~\cite{Hanada:2019rzv}.}
\label{fig:GWW}
\end{figure}

The phase structure of the system is shown in Fig.~\ref{fig:partial-deconfinement-free-YM}.
At $T<T_c$, the system is completely (not partially) confined, i.e.~$M=0$ 
(the blue line in Fig.~\ref{fig:partial-deconfinement-free-YM}),
whereas at $T>T_c$ the system is completely deconfined, i.e.~$M=N$
(the red line in Fig.~\ref{fig:partial-deconfinement-free-YM}).
In the description based on the canonical ensemble (with temperature $T$ varied 
as a controllable parameter),
there is a first order deconfinement phase transition at $T=T_c$.
This first-order transition is a result of the
coexistence of two sectors (the confined and the deconfined sectors) 
in equilibrium.
As is always the case for a first-order transition resulting from 
such a phase equilibrium, 
one can sharpen one's understanding
 by considering the microcanonical ensemble
(with energy $E$ varied as a controllable parameter).~\footnote{
In the case of partial confinement, even if the transition is
not of first order, the confined and deconfined phases
can coexist~\cite{Hanada:2018zxn, Hanada:2019czd, Hanada:2019kue}.
This happens for example if one introduce fundamental matter~\cite{Hanada:2019kue}.
When the transition is not of first order, there is no need to
distinguish the canonical and the microcanonical ensembles.
}
In this description, at $T=T_c$, the ensemble can be parametrized
by $M$ which increases from $0$ to $N$. 
Equivalently, one can choose a parametrization in terms of the
(trace of the) Polyakov loop $P$, which increases from $0$ to $\frac12$ at $T=T_c$.
This parametrization follows naturally from the computation of the effective action
\cite{Sundborg:1999ue,Aharony:2003sx}.
As we shall show below, thermodynamic quantities such
as energy, entropy, and the distribution of the phases of the Polyakov loop,
can be understood from a single relation between $P$ and $M$~\cite{Hanada:2018zxn, Hanada:2019czd},
\begin{align}
P=\frac{M}{2N}.
\end{align}
From the microcanonical viewpoint, therefore,
$T=T_c$, $M=0$ (i.e.~$P=0$) is the transition point 
from complete to partial confinement, 
and $T=T_c$, $M=N$ (i.e.~$P=\frac{1}{2}$) defines 
the transition from 
partial confinement to complete deconfinement.
As we will explain below the latter transition is a GWW transition. 
We denote thermodynamic quantities at this point with the label GWW below.

{\it This two-component picture of the system explains the
thermodynamic properties of the system.}
Energy and entropy are given by 
the sums of those corresponding to the confined and the deconfined sectors.
The former is of order $N^0$ and is negligible compared to the latter, which is proportional to $M^2$,
since the number of excited degrees of freedom is of order $M^2$.
Hence we have
\begin{eqnarray}
E(T=T_c, P;N)
&=& E(T=T_c, P=\frac12; N) \times \left(\frac{M}{N}\right)^2
= E_{\rm GWW} (N) \times |2P|^2
\label{E-vs-P-YM}
\\
S(T=T_c, P;N)
&=&
S(T=T_c, P=\frac12;N)
\times \left(\frac{M}{N}\right)^2
=
S_{\rm GWW}(N)\times |2P|^2
\label{S-vs-P-YM}
\end{eqnarray}
where we ignored the zero-point energy. 
This relation actually holds for the weak-coupling limit of pure Yang-Mills on S$^3$.
From these relations, we obtain 
\begin{eqnarray}
E(T=T_c, P=M/2N,N)
=
E_{\rm GWW}(M), 
\qquad
S(T=T_c, P=M/2N,N)
=
S_{\rm GWW}(M),
\nonumber\\  
\label{eq:energy-entropy-M-YM}
\end{eqnarray}
where $E_{\rm GWW}(M)$ and $S_{\rm GWW}(M)$ are the energy and entropy at the GWW-transition point in the SU($M$) theory. 
By combining it with the one-to-one mapping \eqref{eq:gauge_theory_symmetrize}, we can see that the partially-confined states dominate thermodynamics. 

\begin{figure}[htbp]
\begin{center}
\scalebox{0.3}{
\includegraphics{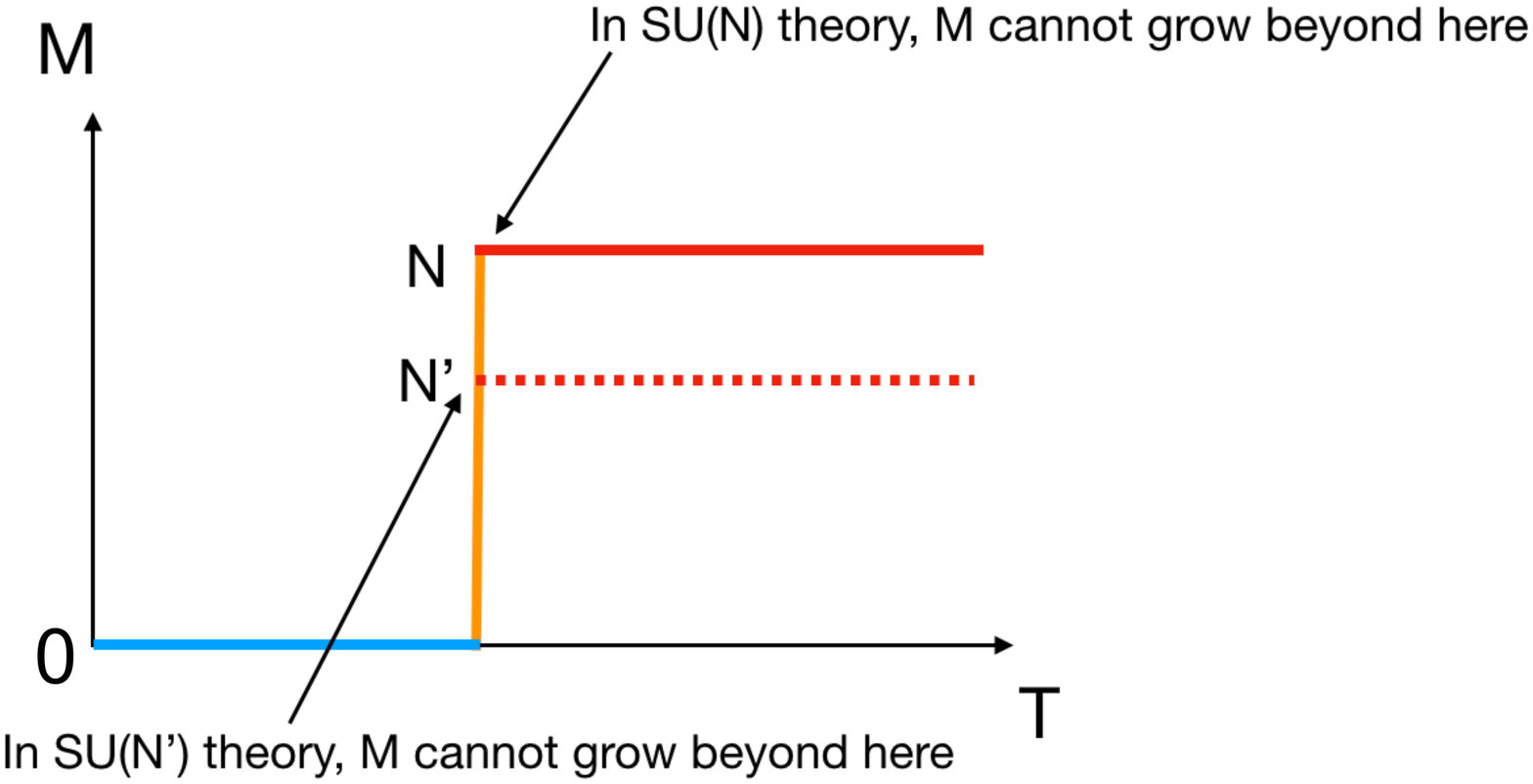}}
\end{center}
\caption{
In the weak-coupling limit,
the deconfined sector at $M=N'$ corresponds to the GWW-transition point of the SU($N'$)-theory. 
This figure is taken from Ref.~\cite{Hanada:2019rzv}. 
}\label{fig:how-to-relate-M-to-GWW}
\end{figure}

The essence of these relations \eqref{eq:energy-entropy-M-YM} is as follows \cite{Hanada:2016pwv,Hanada:2018zxn}. 
Consider SU($N$)- and SU($N'$)-theories, with $N'<N$ (Fig.~\ref{fig:how-to-relate-M-to-GWW}). 
Since energy and entropy are dominated by the deconfined sector, 
there is no apparent difference between the SU($N$)- and SU($N'$)-theories until 
the size of the deconfined sector $M$ reaches $N'$. 
(Note that we are assuming the weak-coupling limit here. At finite coupling, all color degrees of freedom can interact with each other,
and hence the SU($N$)- and SU($N'$)-theories can behave differently.)
Beyond this point, SU($N'$) is completely deconfined, $M$ cannot grow further, and hence the system lies at the GWW-transition point in the SU($N'$)-theory. 
Therefore, the relations \eqref{eq:energy-entropy-M-YM} follow naturally. 

The distribution of Polyakov loop phases can also be clarified from
this point of view.
As shown in Fig.~\ref{fig:GWW}, 
in the completely confined phase $T<T_c$ the distribution $\rho(T; N; \theta)$
is constant. 
For the completely deconfined phase $T>T_c$, $\rho$ becomes $0$ at some 
finite value (i.e., there is a gap). 
At $T=T_c$, for $0<P<\frac12$, the phase distribution is inhomogenous but has no gap.
At $P=\frac12$, a gap forms and 
thus this point is the GWW transition.
The distribution $\rho$ can be computed explicitly by the effective action 
approach~\cite{Sundborg:1999ue,Aharony:2003sx}.
At $T=T_c$, 
it is given by \cite{Sundborg:1999ue,Aharony:2003sx}
\begin{align}
\rho(T=T_c, P;N;\theta)
=&
\frac{1}{2\pi}\left(1+2P\cos\theta\right)
\end{align}
up to $1/N$ corrections. 
By using $P=\frac{M}{2N}$, one 
can rewrite this as
\begin{align}
\rho(T=T_c, P;N;\theta)
=&
 \frac{1}{2\pi}
\left(
1
-
\frac{M}{N}
\right)
+
\frac{M}{N}
\cdot
\frac{1}{2\pi}\left(1+\cos\theta\right)
\nonumber\\
=& \frac{1}{2\pi}
\left(
1
-
\frac{M}{N}
\right)
+
\frac{M}{N}
\cdot
\rho_{\rm GWW}(\theta; M). 
\label{eq:Polyakov_YM}
\end{align}
The first term (which is a constant) and the second term
(which becomes zero at $\theta=\pm\pi$) are the 
respective contributions from the confined and deconfined phases. 
We see that $M$ phases are in the deconfined sector, while the rest is in the confined sector.
Note that $\rho_{\rm GWW}(\theta; M)$ can have a nontrivial $M$-dependence in general (e.g., when the fundamental quarks are added~\cite{Hanada:2019kue}), though for pure Yang-Mills it is simply $\rho_{\rm GWW}(\theta; M)=\frac{1+\cos\theta}{2\pi}$.

In this subsection, we have observed the features 1, 2 and 3 mentioned in the introduction; 
we have studied the free theory and shown that 
among $N^2$ color degrees of freedom, $N^2-M^2$ fall into the confined sector, 
while $M^2$ degrees of freedom are excited.  
Thermodynamic quantities can be understood from the phase equillibrium
between the confined and deconfined sectors.
The feature 4 will be explained in Sec.~\ref{sec:underlying_mechanism} and Sec.~\ref{sec:Pol-vs-ODLRO}.

\subsection{Gauged O($N$) Vector Model at weak coupling}\label{sec:O(N)-vector-model}
\begin{figure}[htbp]
\begin{center}
\scalebox{0.3}{
\includegraphics{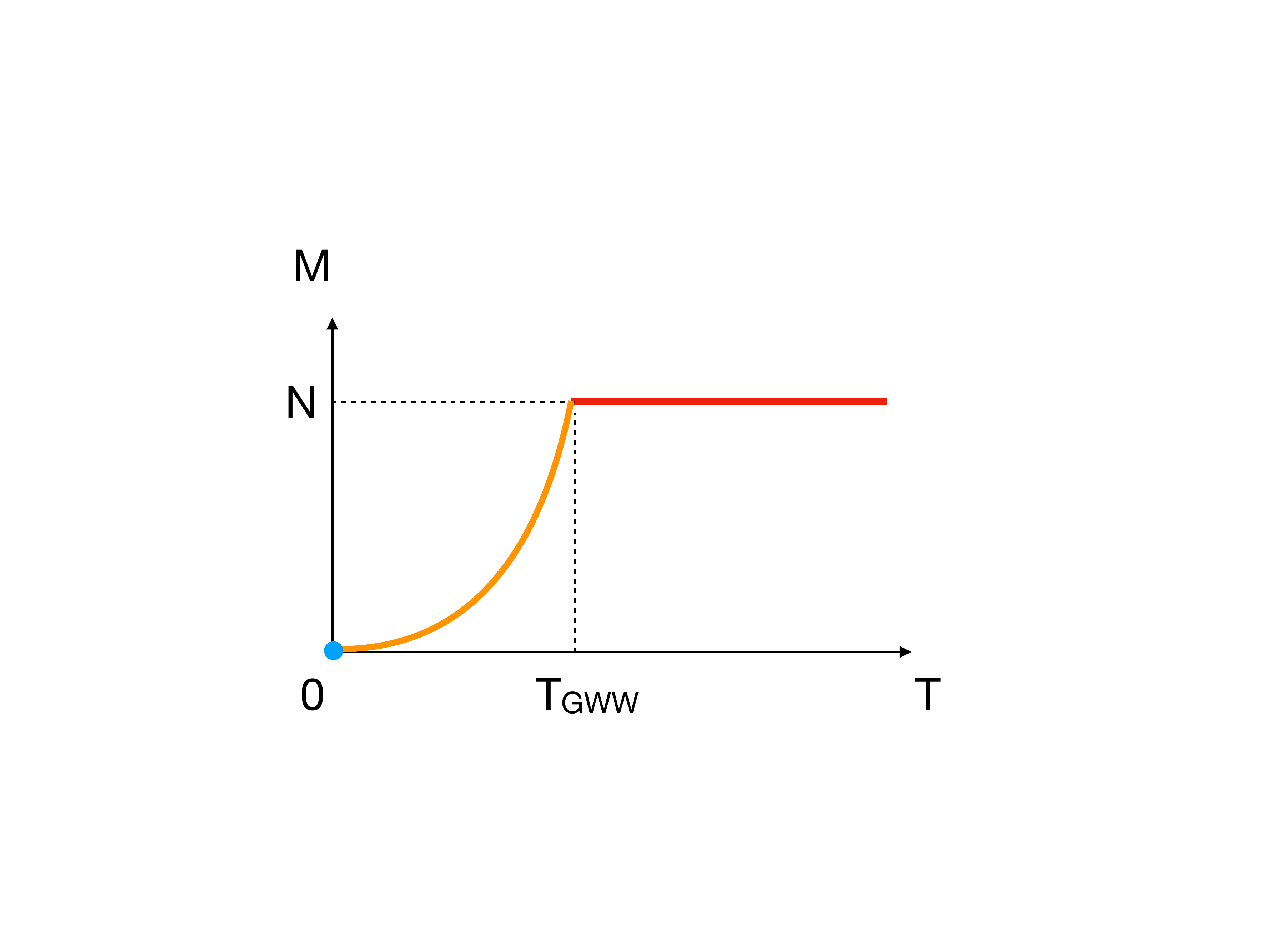}}
\end{center}
\caption{
A schematic picture of the size of the deconfined block $M$ as a function of temperature $T$, 
in the free gauged O($N$) vector model on S$^d$. 
The system is partially confined at $0<T<T_{\rm GWW}$. 
}\label{fig:M-vs-T-vector}
\end{figure}

\hspace{0.51cm}
In this section we consider the gauged O($N$) vector model. 
It is a particularly instructive example to understand the connection between 
confinement at weak coupling and BEC. 
As we will explain in this section, due to the gauge-singlet constraint, this model can exhibit the transition from confinement to deconfinement, characterized by the increase of entropy from $N^0$ to $N^1$ (see also Fig.~\ref{fig:M-vs-T-vector}) \cite{Shenker:2011zf}\footnote{
This scaling holds in an appropriate double scaling limit involving the radius and $N$. 
See the end of this section for details.
}.
As we will see below, 
{\it partial confinement takes place even in the free limit} \cite{Hanada:2019czd}.

We start with the 3d free theory on the two-sphere of radius $R$, 
following Ref.~\cite{Shenker:2011zf}.\footnote{
For simplicity we set the number of flavor $N_f$ in Ref.~\cite{Shenker:2011zf} to be one. }$^,$
We consider an $N$-component vector of scalar fields $\vec{\phi}(x)=\left(\phi_1(x),\cdots,\phi_N(x)\right)$
which transforms in the fundamental representation of the O($N$) symmetry group and consider its free theory
in the O($N$)-singlet sector. 
The $N$ components of $\vec{\phi}(x)$ resemble the $N$ bosons that will be discussed in Sec.~\ref{sec:BEC-identical-bosons}, 
and the gauge symmetry resembles the permutation symmetry.  

In its bare essentials, projection onto the singlet sector is achieved by coupling $\vec\phi$ to a Lagrange multiplier $\lambda$ via
\begin{equation}
	\label{eq:on_action}
	S = \int -\vec{\phi}^T\left(D_{t}^2 + \partial_i^2 - \frac{1}{4}\right)\vec{\phi}\,,
\end{equation}
where the Lagrange multiplier appears inside a gauge covariant derivative $D_t \equiv \partial_t + i\lambda$. 
In three dimensions, this theory appears in the weak coupling or small radius limit of an interacting conformal field theory. More precisely, one can conformally couple $\vec{\phi}(x)$ to an
O($N$) gauge field $A_\mu$ with Chern-Simons action; the free singlet theory is then obtained by taking the level to infinity. The Lagrange multiplier appears as the gauge holonomy around the thermal circle that survives the free limit and is in this way connected to the Polyakov loop. For simplicity of discussion, we will from now on only refer to the latter.

At finite temperature, \eqref{eq:on_action} can be studied in exactly the same way as our previous example of SU($N$) Yang-Mills following the basic idea and tools introduced by \cite{Sundborg:1999ue,Aharony:2003sx}.~\footnote{
	Since the gauge group is SO($N$) rather than U($N$),
	the density $\rho(\theta)$ necessarily becomes symmetric under the refelection 
	$\theta\leftrightarrow -\theta$.
}. Explicitly, one derives an effective action for 
the phase of the Polyakov loop
after integrating out all massive excitations~\cite{Amado:2016pgy,Shenker:2011zf}.
After minimizing the effective action, 
the Polyakov loop is zero at $T=0$, nonzero at any $T>0$, 
and the GWW transition, which is the transition to complete deconfinement, takes place at $T_{\rm GWW}(N)=\frac{\sqrt{3}}{\pi R}\sqrt{N}$. 

Below the GWW-temperature, an energy eigenstate can be expressed in the form of 
eq.~\eqref{eq:gauge_theory_symmetrize}, 
\begin{eqnarray}
|E\rangle_{\rm inv}
=
{\cal N}^{-1/2}\int dU\ {\cal U}\left(|E;{\rm O}(M)\rangle\right). 
\label{eq:vector_model_symmetrize}
\end{eqnarray}  
Here $|E;{\rm O}(M)\rangle$ denotes states for which 
$\phi_{M+1},\cdots,\phi_N$ are in the ground state, 
as shown in the right panel of Fig.~\ref{fig:partial-deconfinement}.
In this way, {\it a large fraction of the degrees of freedom falls into the ground state
}(the confined sector), and the {\it confined sector and deconfined sector coexist}. 

In order to show the above more explicitly and to see
how temperature and $M$ are related, let us look at the Polyakov loop closely. 
By using $b=\frac{TR}{\sqrt{N}}$, and taking $b$ to be of order one, 
the distribution of the Polyakov loop phase $\theta$ is written as
\begin{eqnarray}
\rho(\theta)
=
\frac{1}{2\pi}
+
\frac{2b^2}{\pi}f(\theta), 
\end{eqnarray}
where 
\begin{eqnarray}
f(\theta)
=
-\frac{\pi^2}{12}+\frac{\left(|\theta|-\pi\right)^2}{4}.
\end{eqnarray}
At $b=b_{\rm GWW}=\frac{\sqrt{3}}{\pi}$, the GWW transition takes place; the distribution becomes zero at $\theta=\pm\pi$.

We can rewrite $\rho(\theta)$ as 
\begin{eqnarray}
\rho(\theta)
=
\left(
1
-
\frac{b^2}{b_{\rm GWW}^2}
\right)
\cdot
\rho_{\rm confine}(\theta)
+
\frac{b^2}{b_{\rm GWW}^2}
\cdot
\rho_{\rm GWW}(\theta), 
\end{eqnarray}
where $\rho_{\rm GWW}(\theta)$ is the distribution of the phases at $b=b_{\rm GWW}$, 
and $\rho_{\rm confine}(\theta)=\frac{1}{2\pi}$ is the distribution of the phases in the confined phase. 
The parameter $b$ is related to the size of the deconfined sector $M$ as \cite{Hanada:2019czd}
(see Fig.~\ref{fig:M-vs-T-vector})
\begin{eqnarray}
\frac{M}{N}
=
\frac{b^2}{b_{\rm GWW}^2}.  
\label{eq:T-and-M-1}
\end{eqnarray}
Equivalently, 
\begin{eqnarray}
TR
=
b\sqrt{N}
=
b_{\rm GWW}\sqrt{M}. 
\label{eq:T-and-M-2}
\end{eqnarray}

Note that this is the critical temperature of the O($M$) theory: $T_{\rm GWW}(M)=b_{\rm GWW}\sqrt{M}$. 
Therefore, the identification leads to 
\begin{eqnarray}
\rho(\theta,T=T_{\rm GWW}(M))
=
 \frac{1}{2\pi}
\left(
1
-
\frac{M}{N}
\right)
+
\frac{M}{N}
\cdot
\rho_{\rm GWW}(\theta; M). 
\label{eq:Polyakov_vector_model}
\end{eqnarray}
This relation is analogous to \eqref{eq:Polyakov_YM}. 
As expected, $M$ phases are in the deconfined sector, while the rest is in the confined sector \cite{Hanada:2018zxn,Hanada:2019czd}. 

At $1 \ll T\le T_{\rm GWW}$, the energy scales as $E=AT^5$, 
with an $N$-independent coefficient $A=16\,\zeta(5)$, 
and the entropy is $S=\frac{5}{4}AT^4$.  
Therefore, at $T=T_{\rm GWW}(M)$, 
the following relations hold: 
\begin{eqnarray}
E=E_{\rm GWW}(M), 
\qquad
S=S_{\rm GWW}(M).  
\label{eq:E-and-S-vs-M}
\end{eqnarray}
These equations are analogous to \eqref{eq:energy-entropy-M-YM}. 
Namely, energy and entropy, which are dominated by the deconfined sector, can precisely be explained by O($M$)-partial-confinement. 
As in the case of free Yang-Mills, 
the Polyakov loop, energy and entropy are consistently explained by the same $M$
defined by \eqref{eq:T-and-M-1} and \eqref{eq:T-and-M-2}; 
the O($M$)-deconfined phase  of O($N$)-model
corresponds to the GWW point in the O($M$)-model.
In this way, {\it the two-component picture of the system explains the
thermodynamic properties of the system.}

These calculations can be generalized to $(d+1)$-dimensional theory on S$^d$.  
The critical temperature of the O($N$)-theory is \cite{Amado:2016pgy,Shenker:2011zf}
\begin{eqnarray}
T_{\rm GWW}(N)
=
\left(
\frac{N}{4(1-2^{1-d})\zeta(d)}
\right)^{1/d}\cdot\frac{1}{R},  
\label{sec:vector-model-Tc}
\end{eqnarray}
and the size of the deconfined sector is 
\begin{eqnarray}
\frac{M}{N}
=
\left(
\frac{T}{T_{\rm GWW}}
\right)^d. 
\label{sec:vector-model-M/N}
\end{eqnarray} 
The key relations \eqref{eq:Polyakov_vector_model}\
and \eqref{eq:E-and-S-vs-M} remain unchanged. 

Again, we have seen the features 1, 2 and 3 mentioned in the introduction; 
we have studied the free theory, and shown that 
among $N$ color degrees of freedom, $N-M$ fall into the confined sector, 
while $M$ degrees of freedom are excited.
Once more, feature 4 --- the importance of the interference --- will be explained in Sec.~\ref{sec:underlying_mechanism} and Sec.~\ref{sec:Pol-vs-ODLRO}. 

In Ref.~\cite{Shenker:2011zf}, the radius $R$ is taken to be $N$-independent. 
If instead we take $R=\sqrt{N}$ (or, for generic dimension $d$, $R\propto N^{1/d}$), 
then $b=\frac{TR}{\sqrt{N}}$ simply equals $T$, 
and the natural temperature scale becomes $N$-independent. 
This corresponds to the thermodynamic limit of identical bosons with fixed particle density, 
which is discussed in Sec.~\ref{sec:BEC-identical-bosons}. 
In this limit, the free energy, entropy and energy are of order $N$.
\subsection{BEC of non-interacting particles}\label{sec:BEC-identical-bosons}
\hspace{0.51cm}
The relevance of the BEC of an ideal gas for understanding 
 the superfluidity of $^4$He, which is interacting, 
was first pointed out by F.~London~\cite{1938Natur.141..643L}.  
This idea was elaborated to a two-component fluid theory, corresponding to 
particles in the ground and excited states, respectively, which gave a remarkably good
phenomenological understanding of the superfluidity~\cite{tisza_transport_1938, landau}.
{\it 
The ideal Bose gas (the weak-coupling limit) thus captures a good part of 
the important features of superfluidity (the finite-coupling case).
}
That the introduction of the interaction does not affect these features 
was established through the development of
microscopic understanding of the superfluidity,
in particular through works by 
Feynman, 
Penrose and Onsager~\cite{ feynman_superfluidity1, 
feynman_superfluidity2, feynman_superfluidity3, PhysRev.104.576}.
The validity of Feynman's approach was later confirmed quantitatively
by direct Monte Carlo simulations~\cite{ pollock_simulation_1984, ceperley_path-integral_1986, pollock_path-integral_1987, ceperley_path_1995}.

We will consider in this section the ideal Bose gas 
trapped in a harmonic potential in $d$ spatial dimensions~\cite{deGroot_etal:1950}, 
as the system closely resembles that of the O($N$) vector model studied in the previous 
section.
There are $N$ harmonic oscillators denoted by
$\vec{x}_1,\cdots,\vec{x}_N$, 
each of them having $d$ components. The Hamiltonian is 
\begin{eqnarray}
H
=
\sum_{c=1}^N\left(
\frac{\vec{p}_{c}^{\,\,2}}{2m}
+
\frac{m\omega^2}{2}\vec{x}_{c}^{\,2}
\right).
\end{eqnarray}
\noindent
Because the $N$ particles are indistinguishable bosons, 
invariance under permutations S$_N$ is imposed, which can equivalently be interpreted as gauging the S$_N$ symmetry.
We note that the field $(\phi_1, \cdots, \phi_N)$ of the gauged O($N$) vector model discussed
in the previous subsection is the counterpart of $(\vec{x}_1, \cdots, \vec{x}_N)$ 
in the present model.
Both $x$'s and $\phi$'s belong to the fundamental representation of the
gauge groups S$_N$ and O($N$), respectively.
If we identify the fields $x$ and $\phi$, 
S$_N$ is naturally embedded into O($N$).
In this sense, one may refer to the model considered in this 
section as `S$_N$ vector quantum mechanics'. 
This  close similarity between the gauged O($N$) vector model and 
the system of identical bosons
is what makes the O($N$) model particularly suited to connect the
idea of confinement and BEC.

In the thermodynamic limit of the grand canonical ensemble,
the number of particles
in the excited states $M$ is given by
\begin{eqnarray}
M=
\int_0^\infty d\epsilon \frac{c_d\epsilon^{d-1}}{e^{\beta(\epsilon-\mu)}-1}, 
\end{eqnarray}
where $
c_d
=
\frac{1}{(d-1)!\cdot\omega^d}
=
\frac{1}{\Gamma(d)\cdot\omega^d}$. 
The chemical potential $\mu$ has to satisfy $\mu\le 0$. 
As a function of $\mu$, $M$ is monotonically increasing. 
The largest possible value is given at $\mu=0$. 
Hence, if $M(\mu=0)<N$, a BEC is formed; 
{\it a large fraction of particles, namely $N-M$ of them, are in the ground state.}

These states dominating the condensed phase 
can be written in a form that is analogous to eq.~\eqref{eq:gauge_theory_symmetrize}
and eq.~\eqref{eq:vector_model_symmetrize}. 
To this end, we introduce a set of basis vectors of the system (before imposing the
$S_N$ gauge symmetry, {\it i. e.}, the complete symmetry under the exchanging of particles),
\begin{eqnarray}
|\vec{n}_1,\vec{n}_2,\cdots,\vec{n}_N\rangle
\equiv
\prod_{i=1}^d
\frac{\hat{a}_{i1}^{\dagger n_{i1}}}{\sqrt{n_{i1}!}}
\frac{\hat{a}_{i2}^{\dagger n_{i2}}}{\sqrt{n_{i2}!}}
\cdots
\frac{\hat{a}_{iN}^{\dagger n_{iN}}}{\sqrt{n_{iN}!}}
|0\rangle.
\label{harmonic_oscillator_basis}
\end{eqnarray} 
The state of each particle in the $d$-dimensional harmonic oscillator potential
is specified by a $d$-dimensional interger-valued vector $\vec{n}$, 
where $n_i=0, 1, \cdots$ with $i=1, \cdots, d$.
The $n_{i 1}$ above specifies the state of particle $1$, and so forth.
By using this notation, the states in the condensed phase are,
\begin{align}
P |\vec{n}_1, \cdots, \vec{n}_{M}, \vec{0}, \cdots, \vec{0}\rangle
\end{align}
Here 
$\hat{g}$ is 
the group element $g\in S_N$ 
represented as a unitary operator acting on the Hilbert space.
and 
$\hat{P}=\frac{1}{N!}\sum_{g\in{\rm S}_N}\hat{g}$ is the projection operator 
(the complete symmetrization operator).
The un-symmetrized state $|\vec{n}_1, \cdots, \vec{n}_{M}, \vec{0}, \cdots, \vec{0}\rangle$
is analogous to $|E;{\rm SU}(M)\rangle$ in eq.~\eqref{eq:gauge_theory_symmetrize}
and $|E;{\rm O}(M)\rangle$ in eq.~\eqref{eq:vector_model_symmetrize},  
shown pictorially in Fig.~\ref{fig:partial-deconfinement}. 

In the following, we will denote $M(\mu=0)$ simply by $M$. 
By using 
$
\int_0^\infty du \frac{u^{d-1}}{e^{u}-1}
=
\zeta(d)\Gamma(d)$,
we obtain
\begin{eqnarray}
M
=
\frac{T^{d}\zeta(d)}{\omega^d},   
\end{eqnarray}
and hence the transition temperature $T_c$ is determined by,
\begin{eqnarray}
N
=
\frac{T_c^{d}\zeta(d)}{\omega^d}.   
\end{eqnarray}
Therefore we have
\begin{eqnarray}
\frac{M}{N}
=
\left(
\frac{T}{T_c}
\right)^d
\end{eqnarray} 
and 
\begin{eqnarray}
T_c
=
\left(
\frac{N}{\zeta(d)}
\right)^{1/d}
\omega. 
\end{eqnarray}
Curiously, these formulae are almost identical to the 
corresponding ones for the O($N$) vector model, 
\eqref{sec:vector-model-Tc} and \eqref{sec:vector-model-M/N}.

The energy below $T_c$ is 
\begin{eqnarray}
E
=
\int_0^\infty d\epsilon \frac{c_d\epsilon^{d}}{e^{\beta\epsilon}-1}
=
c_dT^{d+1}\zeta(d+1)\Gamma(d+1). 
\end{eqnarray}
We can easily see that, at $T\le T_c$, 
\begin{eqnarray} 
E(T=T_c(M))=E_c(M).\end{eqnarray}
The entropy satisfies a similar relation, 
\begin{eqnarray} 
S(T=T_c(M))=S_c(M).\end{eqnarray}
Evidently, both energy and entropy are carried solely by excited modes.
These relations are the counterpart of \eqref{eq:E-and-S-vs-M}. 
In this way, 
{\it the thermodynamic properties of the system can be understood by
the two-component (particles in the ground state and in the excited states)
.}

Thus far, we observed a striking similarity between BEC in the system of $N$ identical bosons,
and confinement both in the O$(N)$ vector model and Yang-Mills theory. 
In the next section, we will describe the common mechanism that underlies
BEC and confinement. Along the way, we will introduce
the counterpart of \eqref{eq:Polyakov_vector_model} in BEC.

\section{Common underlying mechanism}\label{sec:underlying_mechanism}
\hspace{0.51cm}
We now proceed to explain the common mechanism 
of color confinement and BEC, which is the origin of 
the remarkable similarity between these phenomena described in Sec.~\ref{sec:weak-coupling}. 

As a result of the constraint on the permutation symmetry, 
the system of bosons favors states where many particles are in the same state.
This well-known enhancement effect, sometimes
denoted as positive interference of the wave functions, is the 
essential mechanism responsible for BEC. 
Although a good part of the explanations that we provide below is well-known 
in the context of BEC,
nonetheless, we decide to present it at a level of detail that exposes 
the connection to partial confinement in gauge theory.

For a system of $N$ indistinguishable bosons, 
 permutation invariance can be incorporated by introducing a projection factor into the partition function as
\begin{eqnarray}
Z
=
\sum_{g\in G}{\rm Tr}\left(
\hat{g} e^{-\beta\hat{H}}
\right),  
\label{eq:partition-function-identical-bosons}
\end{eqnarray}
where $G={\rm S}_N$. 
Here, $\hat{g}$ is the group element $g\in G$ represented 
as a unitary operator acting on the Hilbert space. The inclusion of the projection factor allows for the trace to be taken over the full Hilbert space
and as a complete orthonormal basis, we can use 
$|\vec{n}_1,\vec{n}_2,\cdots,\vec{n}_N\rangle$ defined by eq.~\eqref{harmonic_oscillator_basis}. 

As explained in Appendix~\ref{sec:identical_boson_partition_function}, 
the partition function can equivalently be obtained by summing the contributions from the permutation-invariant states, proportional to $\hat{P}|\vec{n}_1,\cdots,\vec{n}_N\rangle$, 
where $\hat{P}=\frac{1}{N!}\sum_{g\in{\rm S}_N}\hat{g}$ is the projection operator. 
For generic excited states (in which no two particles occupy the same state)
the sum over $g$ in \eqref{eq:partition-function-identical-bosons} 
is used up for making the state completely symmetric.
On the other hand, 
the ground state $|\vec{0},\cdots,\vec{0}\rangle$ is genuinely symmetric, 
even before summing over $g$.
This difference is the foundation of the enhancement effect.
We can write the sum more explicitly as
\begin{eqnarray}
Z
&=&
\sum_{g\in {\rm S}_N}
\sum_{\vec{n}_1,\cdots,\vec{n}_N}
\langle\vec{n}_1,\cdots,\vec{n}_N|
\hat{g} e^{-\beta\hat{H}}
|\vec{n}_1,\cdots,\vec{n}_N\rangle
\nonumber\\
&=&
\sum_{\vec{n}_1,\cdots,\vec{n}_N}
e^{-\beta\left(E_{\vec{n}_1}+\cdots E_{\vec{n}_N}\right)}
\sum_{g\in {\rm S}_N}
\langle\vec{n}_1,\cdots,\vec{n}_N|
\hat{g} 
|\vec{n}_1,\cdots,\vec{n}_N\rangle
\nonumber\\
&=&
\sum_{\vec{n}_1,\cdots,\vec{n}_N}
e^{-\beta\left(E_{\vec{n}_1}+\cdots E_{\vec{n}_N}\right)}
\sum_{g\in {\rm S}_N}
\langle\vec{n}_1,\cdots,\vec{n}_N
|\vec{n}_{g(1)},\cdots,\vec{n}_{g(N)}\rangle. 
\label{eq:partition-function-identical-bosons-2}
\end{eqnarray}
If all $N$ particles are in different states, 
only $g=\textbf{1}$ gives rise to a nonzero contribution. 
On the other hand, if all $N$ particles are in the same state, all $g$'s return the same nonzero contribution, 
leading to an enhancement factor of $N!$ compared to 
the case where we do not impose the gauge singlet constraint
(or equivalently the classical Boltzmann statistics).

One can also think of this enhancement 
as a consequence of redundancy in gauge theories:
configurations connected by a gauge transformation are identified.
Equivalently, when the system of $N$ identical bosons is regarded as an `S$_N$ gauge theory', 
`states' connected by a gauge transformation (i.e.~a permutation) --- 
$|\vec{n}_{1},\cdots,\vec{n}_{N}\rangle$ and $|\vec{n}_{g(1)},\cdots,\vec{n}_{g(N)}\rangle$
--- should be identified. 
Consider for example $|\vec{n},\vec{0},\cdots,\vec{0}\rangle$, 
	$|\vec{0},\vec{n},\vec{0},\cdots,\vec{0}\rangle$, ..., 
	$|\vec{0},\cdots,\vec{0},\vec{n}\rangle$. 
	A priori, these are $N$ different states. Once the S$_N$ symmetry is gauged, they are identified and their statistical weight is reduced to precisely match that of the ground state. Explicitly, one sees that the aforementioned enhancement factors, here $(N-1)!$, combine with the degeneracy factor $N$ to yield an overall coefficient of $N!$. In comparison, the ground state is unique but is enhanced by a factor of $N!$.
	For generic excited states,
	there is an $N!$-fold over-counting which is 
	compensated by the absence of the enhancement factor.
	\footnote{
		More generally, if $M$ particles are excited to different excitation levels
		while $N-M$ particles are in the ground state, 
		classically there are $\frac{N!}{(N-M)!}$ different states, 
		and the enhancement factor is $(N-M)!$.
		Hence the weight in the partition function does not depend on $M$. }
	Thus all gauge invariant states contribute with equal weight. As a result, the relative importance of configurations where many particles occupy the same state is significantly increased, compared to a system of distinguishable particles 
	where the gauge singlet constraint 
	is not imposed. 

This mechanism directly carries over to ordinary gauge theory, where
the gauge-singlet constraint (or equivalently,
the Gauss law constraint) is introduced in the same fashion.
Now, $g$ in \eqref{eq:partition-function-identical-bosons}
is an element of the gauge group, e.g. O$(N)$ or SU($N$),
and the sum is replaced by the invariant integral over the gauge group. 
The group element $g$ now coincides with the Polyakov loop.\footnote{
A simple way to understand this is to consider lattice regularization and take the $A_0=0$ gauge.
The unitary link variable along the time direction $U_t$ connecting the Euclidean time $t$ and $t+a$, where $a$ is the lattice spacing, transforms as $U_t\to\Omega_tU_t\Omega_{t+a}^{-1}$. 
We can use this to set all links to unity, except for the one at $t=0$ which is 
 by definition the Polyakov loop.
}
From this prescription, we see that an enhancement mechanism that is
essentially equivalent to the one in BEC 
also applies to 
the gauge theory.
Namely, 
if all degrees of freedom are in their ground states (the fully confined state),
the integral over the gauge group gives a larger factor
compared to states where degrees of freedom are in different 
excited states (the deconfined state).
This argument applies not just to fields
in the fundamental representation, but also to 
those in the adjoint representation, such as gluons. 
Thus, the confined state, rather than the deconfined state, is favored
as a result of the gauge-singlet constraint.
Note that this argument readily generalizes to the interacting theory, as long as 
{\it the confined sector provides positive interference}. 

Note that this mechanism can work for QFT in any spacetime dimensions. 
The important point is that local gauge symmetry leads to an enhancement factor at each
spatial point. To make the story well-defined by introducing a proper regularization, we can
use the lattice Hamiltonian, e.g. the Kogut-Susskind formulation. At finite lattice spacing and
finite lattice points, it is just a ‘matrix model’ consisting of many matrices (link variables and
site variables). 
Therefore, strictly speaking, $\hat{g}$ is $\hat{g}=\otimes_{\vec{x}}\hat{g}_{\vec{x}}$, where $\hat{g}_{\vec{x}}$ is the group element associated with a point $\vec{x}$. Each $\hat{g}_{\vec{x}}$ can be regarded as the Polyakov loop at point $\vec{x}$.
See Appendix~\ref{sec:symmetrization-example} how the symmetrization over the local gauge transformation leads to the gauge singlets. 

That large-$N$ Yang-Mills theory deconfines at higher temperature is usually 
understood as a consequence of the Hagedorn growth of the density of states,
$\Omega(E)\sim e^{\frac{E}{T_{\rm H}}}$ \cite{Sundborg:1999ue,Aharony:2003sx}
at $E\lesssim N^2$, where $T_{\rm H}$ is the Hagedorn temperature \cite{Hagedorn:1965st}. 
This particular growth rate with respect to the energy allows for a dramatic growth of the energy and entropy as functions of temperature, at $T=T_H$, from order $N^0$ to order $N^2$. At infinite $N$ this is the well-known Hagedorn growth that is obtained by counting the number of singlet states, using the chromoelectric string picture. 
If the singlet constraint were not imposed, the density of states is always ${\cal O}(N^2)$ and no Hagedorn-like growth can be observed.
The mechanism explained in the previous paragraph gives a complimentary understanding 
of the confinement/deconfinement transition. 
Either way one observes the effect of the singlet constraint but from different angles. 

This relationship between confinement and BEC gives us a better understanding of 
why partial confinement occurs 
as depicted in Fig.~\ref{fig:partial-deconfinement}. 
Consider, for example, the possibility that the deconfined sector 
is given by two diagonal blocks whose sizes equal 
$M_1, M_2$ with $M_1^2+M_2^2\approx M^2$, 
while the remaining matrix elements 
are confined. 
Naively, such states would have the same entropy as those shown in Fig.~\ref{fig:partial-deconfinement}, 
because the numbers of excited matrix entries are the same. 
We see now that this type of partial confinement pattern is ruled out
because the volume of the group SU($N-M_1-M_2$) is much smaller than SU($N-M$), 
and hence the enhancement effect is much smaller; therefore these states cannot dominate thermodynamics.  

\section{Polyakov loop and off-diagonal long range order}\label{sec:Pol-vs-ODLRO} 
In the previous sections, we have pointed out that BEC and partial confinement
in large $N$ gauge theories share the essential features listed in the introduction,
based on the discussion in the weak-coupling limit.
In this section, we will consider how our argument can be extended to interacting theories.
For BEC, in the presence of inter-particle interactions, 
Hamiltonian eigenstates are of course no longer given by symmetrized products of individual particle states.
As a consequence, it is not immediately clear how to 
define, for interacting theory, `the number of particles in their ground states'
which characterizes the condensed phase for the ideal gas.
Penrose and Onsager \cite{PhysRev.104.576} 
proposed a criterion valid for interacting theories, 
later referred to as `Off-Diagonal Long Range Order' (ODLRO) \cite{RevModPhys.34.694},
which utilizes 
a natural extension of the concept of `the number of particles in their ground states'. 
For gauge theories, on the other hand, 
the distribution of Polyakov loop phases, as explained in the previous section,
provides a good criterion for  partial confinement, applicable also
to the interacting case \cite{Hanada:2018zxn}.
We will now show that ODLRO in BEC and the Polyakov loop 
in gauge theories are closely related. 
Along the way, we will demonstrate that 
one can define ODLRO for  gauge theories,
and a Polyakov loop for BEC. 
\subsection{Off-diagonal long range order}
We begin by recalling the definition of ODLRO for $N$ identical bosons. 
Denoting the density matrix of the $N$-particle system by $\hat{\rho}$,
the one-particle density matrix is defined by tracing out $N-1$ particles, 
$\hat{\rho}_1=N\cdot{\rm Tr}_{2,3,\cdots,N}\hat{\rho}$. 
It can be conveniently written via its
spectral decomposition, 
\begin{eqnarray}
\hat{\rho}_1
=
n_{\rm max}|\Psi\rangle\langle\Psi|
+
\sum_{i}n_i|\Psi_i\rangle\langle\Psi_i|, 
\label{eq:Penrose-Onsager}
\end{eqnarray}
where $n_{\rm max}$ is the largest eigenvalue and $|\Psi\rangle$ is the corresponding eigenvector. 
The eigenvectors $|\Psi\rangle$ and $|\Psi\rangle_i$ are normalized to be unit norm. 
When $n_{\rm max}$ is of order $N$, the system contains a BEC, 
and is characterized by ODLRO. 
For a BEC of non-interacting bosons, $|\Psi\rangle$ is the one-particle ground state,
and we have $n_{\rm max}=N-M$, i. e.  the number of particles in the ground state.

In the usual thermodynamic limit with fixed particle density, 
$V\sim\omega^{-d}\sim N$, 
the reduced density matrix $\langle x|\hat{\rho}_1|y\rangle$ 
is non-vanishing at long distance if $n_{\rm max}$ is of order $N$. 
The order is associated with the off-diagonal matrix elements in the coordinate representation; 
this is the origin of the name of ODLRO. 

\subsection{Polyakov loop for identical bosons}\label{sec:Polyakov_BEC}
\hspace{0.51cm}
Let us start with the partition function \eqref{eq:partition-function-identical-bosons}. 
Again, a convenient basis is \eqref{harmonic_oscillator_basis}. 
Let $M_{\vec{n}}$ ($\sum_{\vec{n}} M_{\vec{n}}=N$) be the number of particles in 
the state specified by $\vec{n}_i=\vec{n}$. 
A permutation $\{g\in \prod_{\vec{n}}{\rm S}_{M_{\vec{n}}}\}$ leaves the corresponding state invariant and gives rise to a nonzero contribution to \eqref{eq:partition-function-identical-bosons}. 
As we have mentioned, this $g$ is the counterpart of the Polyakov loop in gauge theory. 
The distribution of the phases of this `Polyakov loop' can be obtained 
by calculating the average eigenvalue distribution of $\{g=\{g_{\vec{n}}\}\in \prod_{\vec{n}}{\rm S}_{M_{\vec{n}}}\}$.
At large $N$, we can use the typical values of $M_{\vec{n}}$ realized in the BEC. 

As $M_{\vec{0}}\sim N\to\infty$ (i.e. as the BEC is formed), the average eigenvalue distribution of $g_{\vec{0}}\in {\rm S}_{M_{\vec{0}}}$ becomes uniform. 
To see this, let us note that when $g$ is a cyclic permutation of $k$ elements, 
the eigenvalues of $g$ are $e^{2\pi il/k}$, $l=0,1,\cdots,k-1$. When $k\to\infty$, the phases are distributed uniformly and continuously between $-\pi$ and $+\pi$. 
Any $g_{\vec{0}}\in {\rm S}_{M_{\vec{0}}}$ can be written as a product of cyclic permutations of different sets of elements, 
and as $M_{\vec{0}}\to\infty$, infinitely long cyclic permutations become dominant.\footnote{
Importance of the dominance of long cyclic permutation in understanding
BEC for interacting bosons is first pointed out by Feynman 
in his microscopic theory of superfluidity of ${}^4$He \cite{feynman_superfluidity1}.
The presence of ODLRO when the long cyclic permutation dominates is shown in 
\cite{PhysRev.104.576}.
}
Therefore, $g_{\vec{0}}\in {\rm S}_{M_{\vec{0}}}$ leads to a uniform distribution. 
This is the counterpart of $\frac{1}{2\pi}\left(1-\frac{M}{N}\right)$ in partial deconfinement.  
Thus we have shown that the particles in the ground state (which is measured
by ODLRO) contribute to the constant offset of the Polyakov loop.

In order to complete the proof of equivalence of the constant offset 
to the number of particles in the ground state as measured by ODLRO, 
it remains to be shown that the particles in excited modes do not 
contribute to the constant offset. 
This is somewhat intricate due to the discreteness of 
the permutation group S$_N$ and we defer the reader to Appendix~\ref{appendix:BEC-Polyakov_loop} for a detailed proof. With this in hand, we can directly read off the number of condensed particles from the Polyakov loop.
Such a formulation, based on the Polyakov loop, has the advantage that 
{\it one can infer the existence of positive interference 
from the nonzero constant offset, regardless of the details 
of the interaction.} 
Even at strong coupling, the same quantity characterizes the number of degrees of freedom in the BEC sector.\footnote{
Here an implicit assumption is that excited modes do not contribute to the constant offset,
which may fail when many light degrees of freedom exist. 
} 
 
\subsection{Polyakov loop in gauge theory and ODLRO}
\hspace{0.51cm}
In the case of a gauge theory, the partition function is given by 
\eqref{eq:partition-function-identical-bosons} with $G$ now denoting the gauge group, e.g. O$(N)$ or SU($N$).
As mentioned before, $g$ corresponds to the Polyakov loop.  
The ground state is responsible for the constant distribution, 
because a generic element in O($N$) or SU($N$) gives a uniform distribution 
at large $N$.
Hence we can count the number of degrees of freedom in the confined sector.\footnote{
This was known in several weakly-coupled theories via explicit analytic calculation
\cite{Hanada:2018zxn,Hanada:2019czd,Hanada:2019kue}, 
but there was no concrete justification. 
} 
This argument applies to any large-$N$ gauge theory regardless of the details
of the field content; 
that the distribution of the Polyakov loop phases becomes uniform in the confined phase 
demonstrates the strong positive interference.   
The constant offset (the minimum of the distribution) is related to the size of the deconfined sector via 
\begin{eqnarray}
{\rm The\ constant\ offset}
=
1-\frac{M}{N}. 
\end{eqnarray}
This is the order parameter\footnote{
Polyakov loop is often used as an order parameter to detect the 
spontaneous breaking of the center symmetry. 
Here we are using the Polyakov loop as the order parameter in a different way.
Not only it applies to theories without the center symmetry, 
it is more precise in the sense that it can distinguish three phases: 
completely-confined, partially-confined, and completely deconfined.
}
of the partial confinement. 
The lattice simulations of the bosonic matrix model \cite{Bergner:2019rca,Watanabe:2020ufk}
provide a concrete example at strong coupling. 

Naturally, we can also define a counterpart of ODLRO for gauge theories,
via a reduced `one-color' density matrix. 
For example we can keep only the zero-mode of one of the color degrees of freedom 
(say the first component of the matter field in the fundamental representation, 
or $(1,1)$-component of the adjoint field) and 
trace out all other degrees of freedom.
The existence of the confined phase can then be read off from
the largest eigenvalue of the reduced density matrix\footnote{
This can be done in a gauge invariant manner, in the same way as the one-particle reduced density matrix is permutation invariant in the case of identical bosons. The density matrix itself 
$\rho= \sum |\Psi_i\rangle e^{-\beta E_i} \langle \Psi_i|$ is 
gauge invariant,  $|\Psi_i\rangle$ satisfying the Gauss law constraint.
Because of this there is no ambiguity from the choice of gauge when 
defining the eigenvalues of the reduced density matrix.
}.
If we normalize the largest eigenvalue of the reduced density matrix in such a way 
that it equals unity for the fully confined (condensed) phase, then it corresponds to 
the constant offset of the Polyakov loop.
We note that ``the long range order'' in this context is longe range 
not in the spacetime but in 
the `emergent space' described by the values of the field. 

We expect that the large positive interference responsible for the constant offset
survives when the interaction is turned on adiabatically, just like ODLRO does. 
Both order parameters (the eigenvalue in ODLRO, and the constant offset 
of the Polyakov loop) are tied to the gauge symmetry as explained in 
Sec.~\ref{sec:underlying_mechanism}.
Because of this, we expect that
for any value of the coupling constant
the two transitition points, namely, from completely-confined to partially-confined phase, and
from partially-confined to the completey-deconfined phase, should be captured by the conditions
that the order parameters be equal to $0$ and $1$, respectively.

\section{Discussions}\label{sec:discussions}
\hspace{0.51cm}

In this paper, we pointed out that two important phenomena, BEC and (partial) confinement,
can be understood in a unified way. 
We expect that, 
because of this new connection, 
computational tools, and perhaps more importantly intuition
developed for one of them 
can now enrich the understanding of the other.
For example, in superfluidity, 
transport properties are well understood in terms of 
a two fluid model corresponding to condensed and excited states;
can we obtain a similar understanding for the transport properties in a
partially confined phase? 

We have focused on model-independent features, such
as the mechanism behind the phenomenon and its essential characterization.
Confinement (condensation) occurs because a large fraction of 
the degrees of freedom fall into the ground state. 
This phase is favored because of the large interference effect originating 
in the gauge symmetry.
More detailed features, such as the precise structure of the phase diagram
(including the existence of a completely condensed phase)
and the order of the phase transition, 
depend on model specifics.~\footnote{
A classic example of this type of model dependence  is the difference between 
the superfluidity of ${}^4$He and the condensation of an ideal Bose gas.
For the ideal gas the transition is of third order, whereas the $\lambda$-transition of
${}^4$He is of second order. The ideal bose gas is completely condensed at $T=0$, whereas
${}^4$He is not. Nevertheless, they share
common characterization (such as ODLRO) and mechanism, and
the analogy to BEC of the ideal Bose gas was an important step to understand  superfluidity. }

Our strategy has been to understand confinement by 
an adiabatic continuation of the weak-coupling (small volume) picture.
Whether this picture remains relevant at strong coupling (large volume) depends 
on the dynamics of the model and in particular relies on the absence of
a phase boundary that obstructs the interpolation 
between strong- and weak-coupling regions.
However, there are indications that such an obstruction is absent 
for various theories important for gauge/gravity duality, most notably 4d ${\cal N}=4$ super Yang-Mills, 
although thus far there exists no direct proof.\footnote{
On the other hand, QCD with too many flavors is conformal at infrared, which suggests the strong dynamics spoils the weak-coupling picture in this case. 
}
Whether the strong and weak-coupling regimes are smoothly connected for a given model
is a question which can be tested by lattice Monte Carlo simulations. 
For the D0-brane quantum mechanics and its plane wave deformation,
extensive numerical studies have been performed, starting from
Refs.~\cite{Anagnostopoulos:2007fw,Catterall:2008yz}, which support the absence of the 
obstruction.

The analogy to BEC also provided new insight on order parameters which 
should be useful to interpolate 
between
the weak and the strong-coupling regimes.
We showed that the constant offset of the distribution of the 
Polyakov loop phases corresponds to ODLRO, and is tied to 
the structure of the gauge symmetry associated with the condensed phase. 
This gives in particular a characterization of partial confinement 
which is valid even at nonzero coupling.
The constant offset $\frac{N-M}{N}$ is the order parameter that encodes 
the size of the deconfined sector: an SU($M$)-subsector of SU($N$)-theory is deconfined.

\subsubsection*{Implication for gauge theories; Connection to QCD?}
\hspace{0.51cm}
In BEC for interacting bosons, `the number of particles in the ground state' 
as defined by ODLRO is less than $N$ even at zero temperature, 
which is in marked contrast with the ideal Bose gas
where for $T=0$ all particles are in the ground state. 
An intriguing possibility is that a similar phenomenon may occur 
for some gauge theories: for these theories
there may not be complete confinement even for $T=0$.

It is interesting to study the connection of 
our understanding of confinement as BEC to the more traditional 
pictures of dynamical confinement (e.g.~based on the linear potential between quarks).
It might be possible to achieve this  through the idea of magnetic monopole condensation
\cite{Nambu:1974zg,Mandelstam:1974pi,tHooft:1981bkw,Seiberg:1994rs} 
which is a promising scenario for dynamical confinement.
In some versions of this scenario, singularities 
plays an important role that occurs when the nature of the degrees of 
freedom associated with the monopole changes 
(e. g. when the monopole becomes massless)~\cite{tHooft:1981bkw,Seiberg:1994rs}. 
What happens at these singularities resembles the
enhancement effect of confined states in our scenario because
of the large interference effect.
Namely, in the partition function \eqref{eq:partition-function-identical-bosons},
while the confined sector is genuinely SU($N-M$)-symmetric, 
the deconfined sector is SU($M$)-symmetric only due to symmetrization.  
In other words, the confined sector is statistically enhanced (positive interference)
while the deconfined sector is not 
(see also Appendix~\ref{sec:identical_boson_partition_function}). 

Given that theories at small volume and large $N$ are often quantitatively close to 
those at large volume and moderate $N$ \cite{Eguchi:1982nm,GonzalezArroyo:1982hz,Parisi:1982gp,Bhanot:1982sh,Gross:1982at}, 
it seems imaginable to also interpret confinement at finite $N$ as BEC.
For example, in the Twisted Eguchi-Kawai reduction \cite{GonzalezArroyo:1982hz}, 
large-$N$ theory at small volume behaves similar to the finite-$N$ theory at volume $V\sim N^2$. 
Closely related phenomena have been studied extensively in QCD-like theories with adjoint fermion \cite{Kovtun:2007py}
or certain deformation terms \cite{Unsal:2008ch}. Such theories would provide us with analytically controllable setups.
Recall that for indistinguishable bosons in a harmonic trap, 
the thermodynamic limit $V\sim\omega^{-d}\sim N$ 
is typically taken with fixed particle density. 
In this limit, interference effects contribute to the free energy with a relative factor 
$\log(N!)\sim V(\log V-1)$. 
In gauge theory, because the gauge group can act locally, 
even when $N$ is fixed there is a similar factor $\sim V\log V_G$, 
where $V_G$ is the volume of gauge group $G$.  
One should be able to understand confinement for finite $N$ gauge theories
as the result of a mechanism similar to that discussed in Sec.~\ref{sec:underlying_mechanism},
because of this large enhancement factor.

Finally, it will be an important step to investigate possible experimental signals in colliders that could indicate whether confinement in actual QCD bears any resemblance with BEC.
\subsubsection*{Condensation of D-branes?}
\hspace{0.51cm}
D-branes play essential roles in string theory. 
As is well-known, their low-energy effective theory 
is a certain Yang-Mills theory coupled to adjoint matter fields with a U($N$) gauge group
\cite{Witten:1995ex}.
Diagonal elements of the adjoint scalar fields corresponds to the location of D-branes.
This U($N$) group contains S$_N$ subgroup which permutes D-branes.
In this sense, U($N$) gauge symmetry can be interpreted as generalization of 
S$_N$ permutation symmetry. 
Consider now the system of D-branes at very low temperature  
such that the typical distance between them is smaller than their
thermal de Broglie wavelength.
In this regime, it is natural to expect
that the D-branes would undergo a quantum statistical transition,
analogous to BEC.\footnote{
Although D-branes are so-called superparticles that 
can be bosonic or fermionic depending on the excitation of their internal degrees of freedom,
the bosonic degrees of freedom will dominate
for low temperature physics we are interested in.
This is because states associated with the fermionic degrees of freedom 
inevitably have much higher energy than their bosonic counterparts, since they live on the Fermi surface due to the Pauli principle.
}
The similarity of  partial confinement to BEC advocated in this paper
makes it  plausible that partial confinement should be crucial in the understanding
of this quantum condensation of D-branes.

\subsubsection*{Holographic emergent space? }
\hspace{0.51cm}
In the standard interpretation, the completely deconfined and confined phases
correspond to the AdS vacuum and a black hole, respectively~\cite{Witten:1998zw}.
A natural candidate for a dual gravity interpretation of partially deconfined and confined sectors
are the small black hole and its exterior \cite{Hanada:2016pwv,Hanada:2018zxn,Hanada:2019czd}. 
According to the analogy to BEC, the small black hole would correspond
to a droplet of normal fluid within superfluid. 
The Hawking radiation then will be analogous to the dissipation of
this droplet.

In the case of four-dimensional ${\cal N}=4$ super Yang-Mills, 
the six scalar fields can condense. Such a BEC is effectively six-dimensional at each point in 3d space, 
thus leading to nine-dimensional space.  One may speculate that
gravity can be understood as collective excitations analogous to phonons in superfluid helium. 
Such an interpretation would provide us with a natural generalization of the philosophy of the Matrix Model of M-theory (BFSS) \cite{Banks:1996vh} --- physical objects are realized as sub-matrices --- to gauge/gravity duality \`{a} la Maldacena. 
Note also that partial deconfinement is naturally connected to Higgsing; when a deconfined block is far separated (in the sense of eigenvalues),
Higgsing is a better description because the off-diagonal elements become heavy and decouple from the dynamics.\footnote{
It is well-known that, in presence of scalar matter fields,
the confinement phase is smoothly connected to the Higgs phase~\cite{Fradkin:1978dv}.} 
When the partially-deconfined sector represents a D-brane probe, it should be described by the Dirac-Born-Infeld action on AdS$_5\times$S$^5$, as proposed in Ref.~\cite{Maldacena:1997re}.
Furthermore note that the color degrees of freedom in the confined sector can be entangled
and naturally lead to a picture for emergent space \cite{Alet:2020ehp,Hanada:2019czd}
along the lines of Refs.~\cite{Maldacena:2001kr,VanRaamsdonk:2010pw}. 
When colors are identified with qubits, `it from qubit' naturally meets the good old idea of `everything from matrices'.  
One may hope that the intuition gained by connecting BEC and confinement will
be a useful guide towards understanding the nature of the building blocks of
emergent spacetime.

\begin{center}
\section*{Acknowledgement}
\end{center}
\hspace{0.51cm}
We would like to thank Yoichi Kazama for critical comments 
and useful discussions. 
We also thank Ofer Aharony, Sinya Aoki, Andy O'Bannon, Brandon Robinson, Andreas Schmitt, Kostas Skenderis, Bo Sundborg, and Naoki Yamamoto for stimulating discussions
and
 Etsuko Itou, Antal Jevicki,
Paul Romatschke, and Stephen Shenker for carefully reading the draft.  
MH  was supported by the STFC Ernest Rutherford Grant ST/R003599/1
and JSPS  KAKENHI  Grants17K1428. NW acknowledges
support by FNU grant number DFF-6108-00340.

\appendix

\section{Another look at positive interference}\label{sec:identical_boson_partition_function}
\hspace{0.51cm}
In order to understand positive interference further, let us see how the partition function 
\eqref{eq:partition-function-identical-bosons-2} for $N$ free bosons is obtained by summing the contribution of permutation-invariant states. 
By using the projection operator $\hat{P}=\frac{1}{N!}\sum_{g\in{\rm S}_N}\hat{g}$, we can write the invariant states as 
\begin{eqnarray}
c^{{-1}}_{\vec{n}_1,\cdots,\vec{n}_N}
\times
\hat{P}|\vec{n}_1,\cdots,\vec{n}_N\rangle.
\label{eq:normalized-symmetrized-states-boson}
\end{eqnarray}
where $c_{\vec{n}_1,\cdots,\vec{n}_N}$ ensures unit normalization. 
For example, when $N=2$, $c_{\vec{n}_1,\vec{n}_2}=1$ for $\vec{n}_1=\vec{n}_2$
and $c_{\vec{n}_1,\vec{n}_2}=\frac{1}{\sqrt{2}}$ for $\vec{n}_1\neq\vec{n}_2$. 
In general, if there are $l$ different one-particle states with degeneracies $N_1,\cdots,N_l$ 
($N_1+\cdots+N_l=N$, $N_i\ge 1$, $l<N$), 
\begin{eqnarray}
c_{\vec{n}_1,\cdots,\vec{n}_N}
=
\sqrt{
	\frac{\prod_{i=1}^l N_i!}{N!}}. 
\end{eqnarray}
Note that 
this factor $\prod_{i=1}^l N_i!$
is related to 
positive interference.

When we calculate the partition function, 
if we took the sum with respect to any $\vec{n}_1,\cdots,\vec{n}_N$, 
we would be counting the same state multiple times, with the over-counting factor {$\frac{N!}{\prod_{i=1}^l N_i!}$}. 
By compensating this factor, we obtain
\begin{eqnarray}
\lefteqn{
	\sum_{\vec{n}_1,\cdots,\vec{n}_N}
	\left({\frac{N!}{\prod_{i=1}^l N_i!}}\right)^{-1}
	\cdot
	c_{\vec{n}_1,\cdots,\vec{n}_N}^{{-2}}
	\cdot
	\langle\vec{n}_1,\cdots,\vec{n}_N|\hat{P}
	e^{-\beta\hat{H}}\hat{P}|\vec{n}_1,\cdots,\vec{n}_N\rangle
}\nonumber\\
&=&
\sum_{\vec{n}_1,\cdots,\vec{n}_N}
\langle\vec{n}_1,\cdots,\vec{n}_N|\hat{P}
e^{-\beta\hat{H}}\hat{P}|\vec{n}_1,\cdots,\vec{n}_N\rangle
\nonumber\\
&=&
\sum_{\vec{n}_1,\cdots,\vec{n}_N}
\langle\vec{n}_1,\cdots,\vec{n}_N|\hat{P}
e^{-\beta\hat{H}}|\vec{n}_1,\cdots,\vec{n}_N\rangle
\nonumber\\
&=&
{\frac{1}{N!}}
\sum_{\vec{n}_1,\cdots,\vec{n}_N}
\sum_{g\in{\rm S}_N}\langle\vec{n}_1,\cdots,\vec{n}_N|\hat{g}
e^{-\beta\hat{H}}|\vec{n}_1,\cdots,\vec{n}_N\rangle.
\end{eqnarray}
This is eq.~\eqref{eq:partition-function-identical-bosons-2}, up to the overall factor {$(N!)^{-1}$}.  
For gauge theory, the symmetrization defined by eq.~\eqref{eq:gauge_theory_symmetrize} 
does exactly the same job:
the symmetrized state in eq.~\eqref{eq:gauge_theory_symmetrize} 
is the counterpart of 
\eqref{eq:normalized-symmetrized-states-boson}.
In Sec.~\ref{sec:Yang-Mills}, 
we started with the SU($M$)$\times$SU($N-M$)-invariant state 
$|E;{\rm SU}(M)\rangle$. That approach is advantageous for the computation of the
entropy. 
However, one can start with a state
without imposing the Gauss law constraint associated with SU($M$);
The symmetrization ~\eqref{eq:gauge_theory_symmetrize} will
assure the SU($M$)-invariance in the deconfined sector.
The deconfined sector is SU($M$)-invariant due to the symmetrization, 
in the same way that the excited sector of the system of identical bosons is S$_M$-invariant.  
In contrast, the confined sector is `genuinely' gauge-invariant, 
even without symmetrization,  
and hence, the enhancement factor, which is the volume of SU($N-M$), appears.

\section{More on ODLRO and Polyakov loop}\label{appendix:BEC-Polyakov_loop}
\hspace{0.51cm}
In this appendix, we prove the equivalence of ODLRO and
the criteria based on the Polyakov loop for the ideal gas in a harmonic 
oscillator potential well by showing 
that excited modes do not contribute to the constant offset of the Polyakov loop. 

For this purpose, it is convenient to express the partition function in the following form.
Suppose that a given element $g\in {\rm S}_N$ is a product of cyclic
permutations with length $l_1,l_2,\cdots$.  
Then, 
\begin{eqnarray}
{\rm Tr}\left(
\hat{g} e^{-\beta\hat{H}}
\right) 
=
\prod_i
\left(
\frac{1}{1-e^{-l_i\beta\omega}}
\right)^d. 
\end{eqnarray}
Let $N_l$ be the number of cyclic permutations with length $l$. 
Then the partition function \eqref{eq:partition-function-identical-bosons} is written as 
\begin{eqnarray}
Z
=
\sum_{\{N_l\}}
\frac{N!}{\prod_l\left(l^{N_l}\cdot N_l!\right)}
\left(
\frac{1}{1-e^{-l\beta\omega}}
\right)^{dN_l}. 
\end{eqnarray}
Here the sum is taken over all possible $\{N_l\}$ satisfying $\sum_l lN_l=N$. 
We introduce a Lagrange multiplier (chemical potential) $\mu$ to enforce this constraint 
and minimize the free energy to obtain 
\begin{eqnarray}
N_l=\frac{e^{-\mu l}}{l(1-e^{-l\beta\omega})^d}, 
\qquad
\sum_{l=1}^N lN_l=N. 
\label{RFNl}
\end{eqnarray}
In the large $N$ limit, the Polyakov loop $g$ specified with these $N_l$'s dominates.

The eigenvalue distribution of $\hat{g}$ is determined by $N_l$ as explained in the main text.
Namely, for each cyclic permutation with length $l$, there are $l$ numbers of eigenvalues,
$e^{2\pi ik/l}$, $k=0,1,\cdots,l-1$.
The total number of the eigenvalues is, of course, $\sum l N_l =N$.
Our task is to understand the distribution function $\rho(\theta)$ 
of the phases of the eigenvalues determined by \eqref{RFNl} in the large $N$ limit.
We choose $\theta$ to be in the range $0\le \theta < 2\pi$ for convenience and 
normalize $\rho(\theta)$ by $\int \rho d\theta =1$.

We will first consider the case $T=T_c$.
We denote $\rho(\theta)$ in this critical case as $\rho_c(\theta)$. 
The statement we wish to prove first is that the constant offset, i.e. the minumum of 
$\rho_c(\theta)$, vanishes. Note that $\rho_c(\theta)\ge 0$ by definition.
Our strategy is as follows: 
we will write
the distribution $\rho_c(\theta)$ as a sum of two terms, namely, contributions from 
$l<\Lambda$ and $l \ge \Lambda$.
\begin{align}
\rho_c(\theta)=
\rho_{c, l<\Lambda}(\theta) +
\rho_{c, l\ge\Lambda}(\theta)
\end{align}
where $\Lambda$ is a large 'cutoff'.
By definition $\rho_{c, l<\Lambda}(\theta)\ge 0$ and $\rho_{c, l\ge\Lambda}(\theta)\ge 0$.
One may imagine that 
we are evaluating $\rho_c( \theta)$ as $\rho_{c, l<\Lambda}(\theta)$ up to a certain precision,
or equivalently a finite resolution in $\theta$, 
of $\delta\theta\sim\frac{2\pi}{\Lambda}$.
The larger the value of $\Lambda$ the more precise our evaluation of $\rho_c(\theta)$ will be.
We will show that the minimum of $\rho_{c, l<\Lambda}(\theta)$ vanishes for any finite $\Lambda$,
and $\rho_{c, l\ge\Lambda}(\theta)$ (and hence its constant offset) can be made arbitrarily small
by choosing $\Lambda$ to be sufficiently large but of order $N^0$.

At $T=T_c$, since $\mu=0$ and $\beta\omega=\left(\frac{\zeta(d)}{N}\right)^{\frac{1}{d}}$,
the formula \eqref{RFNl} yields
\begin{eqnarray}
\lim_{N\to\infty}\frac{\sum_{l<\Lambda} lN_l}{N}
=
1-\sum_{l=\Lambda}^\infty\frac{1}{\zeta(d)l^d},
\end{eqnarray}
for large $\Lambda$.
It is essential that for $d>1$, which is the condition for BEC to occur, the
second term converges and is of order 
$O(\Lambda^{-(d-1)})$.
For $\Lambda \sim N^0$, $\rho_{l<\Lambda}(\theta)$ is a sum of 
finite (i. e. of order $N^0$) number of delta functions (located 
at $2\pi k/l$, $k=0, \cdots, l-1$ where $l< \Lambda$). 
The minimum of $\rho_{c, l<\Lambda}(\theta)$ is zero, for any finite value of $\Lambda$.
On the other hand
$\rho_{c, l\ge\Lambda}(\theta)$ may approach a continuous function in the large $N$ limit.
In particular $\rho_{c, l\ge\Lambda}(\theta)$ could have contributed to 
a constant offset for $N\to \infty$. 
However,
since 
\begin{align}
\int \rho_{c, l\ge\Lambda}(\theta) d\theta =
\sum_{l\ge \Lambda} \frac{l N_l}{N}
\approx
\sum_{l\ge \Lambda}\frac{1}{\zeta(d)l^d}=O(\Lambda^{-(d-1)})
\end{align}
for large $N$, the function $\rho_{c, l\ge\Lambda}(\theta)$ itself 
can be made arbitrarily small by choosing 
sufficiently large (but of order $N^0$) $\Lambda$.
Thus we have shown $\rho_c(\theta)$ have a vanishing constant offset 
in the large $N$ limit.

For $T\le T_c(N)$, the BEC is formed. 
The statistical distribution of the excited particles (and therefore the contribution
to $\rho(\theta)$ from the excited states) is 
identical to that for the system with $N=M$ if we fix $M$ by $T=T_c(M)$.
The ground state contributes a constant term,
$\frac{1}{2\pi}
\left(
1-\frac{M}{N}
\right)$, as explained in the main text.
Therefore, 
we obtain 
\begin{eqnarray}
\rho(\theta)
=
\frac{1}{2\pi}
\left(
1-\frac{M}{N}
\right)
+
\frac{M}{N}\rho_{c, l < \Lambda} (\theta)
+
O(\Lambda^{-(d-1)}),
\end{eqnarray}
which may be considered as the counterpart of \eqref{eq:Polyakov_vector_model}
in partial deconfinement. Again one can choose $\Lambda$ to be sufficiently large 
(but of order $N^0$) such that the last term is negligible.
The formula shows that the contribution of the constant offset is solely from the ground state, 
which completes the proof.

\section{More on the gauge-invariant states via the symmetrization}\label{sec:symmetrization-example}
In this appendix we consider a gauge-invariant operator constructed using the Wilson line,
\begin{align}
\sum_{i,j=1}^N\hat{q}_i(x)^\dagger\hat{W}_{ij} (x, y)\hat{q}_j(y)
\end{align}
where $\hat{q}$ represents a quark field and $\hat{W}(x, y)$ is a Wilson line connecting points $x$ and $y$.
We show that this is obtained via the symmetrization over the gauge symmery. 
Specifically, let us see how this gauge-invariant combination is obtained from 
$\hat{q}_I(x)^\dagger\hat{W}_{IJ} (x, y)\hat{q}_J(y)$, where in the latter the sums over $I$ and $J$ are not taken. 
Firstly we symmetrize over the U($N$) symmetry at point $x$. The SU($N$) transformation is given by 
\begin{align}
\hat{q}_I(x)^\dagger
\to
\sum_{i=1}^N\hat{q}_i(x)^\dagger U^\dagger_{iI}(x), 
\end{align}
\begin{align}
\hat{W}_{IJ} (x, y)
\to
\sum_{i'=1}^NU_{Ii'}(x)\hat{W}_{i'J} (x, y), 
\end{align}
and
\begin{align}
\hat{q}_I(x)^\dagger \hat{W}_{IJ} (x, y)\hat{q}_J(y)
\to
\sum_{i,i'}\hat{q}_i(x)^\dagger U^\dagger_{iI}(x)U_{Ii'}(x) \hat{W}_{i'J} (x, y)\hat{q}_J(y). 
\end{align}
After averaging over the Haar measure, $U^\dagger_{iI}(x)U_{Ii'}(x)$ is replaced with $\delta_{ii'}$. 
Hence the symmetrization over SU($N$) at point $x$ leads to
\begin{align}
\hat{q}_I(x)^\dagger \hat{W}_{IJ} (x, y)\hat{q}_J(y)
\to
\sum_{i=1}^N\hat{q}_i(x)^\dagger \hat{W}_{iJ} (x, y)\hat{q}_J(y). 
\end{align}
We can perform the symmetrization over SU($N$) at point $y$ as well, and obtain: 
\begin{align}
\sum_{i=1}^N\hat{q}_i(x)^\dagger \hat{W}_{iJ} (x, y)\hat{q}_J(y) 
\to
\sum_{i=1}^N\sum_{j=1}^N\hat{q}_i(x)^\dagger \hat{W}_{ij} (x, y)\hat{q}_j(y). 
\end{align}

In the same manner, various gauge-invariant operators, both local and nonlocal, are obtained via the symmetrization.

\bibliographystyle{utphys}
\bibliography{BEC-Confinement}

\end{document}